\documentclass[hyper,12pt,letterpaper]{JHEP3}

\usepackage{latexsym,amsmath,amsfonts,amssymb}
\usepackage{mathrsfs}
\usepackage[makeroom]{cancel}
\usepackage{bbm}
\usepackage{bm}
\usepackage{subfigure}
\usepackage{paralist}
\usepackage{url}

\numberwithin{equation}{section}

\newcommand{\mat}[1]{\begin{pmatrix} #1 \end{pmatrix}}

\newcommand{\be}{\begin{equation}} 
\newcommand{\ee}{\end{equation}}
\newcommand{\bea}{\begin{equation} \begin{aligned}} \newcommand{\eea}{\end{aligned} \end{equation}}

\newcommand{\bit}{\begin{itemize}} 
\newcommand{\eit}{\end{itemize}}

\newcommand{\cI}{\mathcal{I}}

\newcommand{\cZ}{\mathcal{Z}}

\newcommand{\bC}{\mathbb{C}}

\newcommand{\bZ}{\mathbb{Z}}

\newcommand{\Z}{\mathbb{Z}}
\newcommand{\C}{\mathbb{C}}
\newcommand{\R}{\mathbb{R}}

\renewcommand{\t}{\widetilde }
\renewcommand{\d}{\partial }

\renewcommand{\b}{\bar }

\newcommand{\half}{{1\over 2}}

\newcommand{\bz}{{\b z}}

\newcommand{\CB}{\mathcal{B}}
\newcommand{\CC}{\mathcal{C}}

\newcommand{\CE}{\mathcal{E}}
\newcommand{\CF}{\mathcal{F}}
\newcommand{\CG}{\mathcal{G}}
\newcommand{\CH}{\mathcal{H}}

\newcommand{\CJ}{\mathcal{J}}

\newcommand{\CL}{\mathcal{L}}

\newcommand{\CN}{\mathcal{N}}
\newcommand{\CO}{\mathcal{O}}

\newcommand{\CQ}{\mathcal{Q}}
\newcommand{\CR}{\mathcal{R}}
\newcommand{\CS}{\mathcal{S}}
\newcommand{\CT}{\mathcal{T}}

\newcommand{\CV}{\mathcal{V}}

\newcommand{\CY}{\mathcal{Y}}
\newcommand{\CZ}{\mathcal{Z}}

\newcommand{\FR}{\mathfrak{R}}
\newcommand{\Fg}{\mathfrak{g}}
\newcommand{\Fh}{\mathfrak{h}}

\newcommand{\GG}{\mathbf{G}}

\newcommand{\GH}{\mathbf{H}}

\newcommand{\rk}{{{\rm rk}(\GG)}}

\newcommand{\h}{\hat}

\newcommand{\oneloop}{\text{1-loop}}

\DeclareMathOperator{\Tr}{Tr}
\DeclareMathOperator{\tr}{tr}

\DeclareMathOperator{\sign}{sign}

\newcommand{\rs}{{\bf r}}

\newcommand{\SL}{{\mathscr L}}

\newcommand{\p}{\partial}

\newcommand{\ov}{\over}

\newcommand{\ra}{\rightarrow}

\newcommand{\hD}{{\hat D}}
\newcommand{\eff}{\text{eff}}

\newcommand{\fM}{\mathfrak{M}}
\newcommand{\tfM}{{\widetilde{\mathfrak{M}}}}

\newcommand{\setcond}[2]{\{\,#1\,:\,#2\,\}}

\newcommand{\tp}{\theta^+}

\newcommand{\ttp}{{\t\theta}^+}

\title{
Localization of twisted $\mathcal{N}{=}(0,2)$ gauged linear sigma models in two dimensions 
}

\author{Cyril~Closset,$^1$   Wei Gu,$^2$ Bei Jia$^3$ and Eric Sharpe$^2$\\

{}$^{1}$ Simons Center for Geometry and Physics\\ 
 State University of New York,  Stony Brook, NY 11794, USA\\

{}$^{2}$ Department of Physics MC 0435, 850 West Campus Drive  \\
 Virginia Tech,  Blacksburg, VA 24061, USA\\

{}$^{3}$  Theory Group, Physics Department \\
 University of Texas, Austin, TX 78612, USA

}

\preprint{UTTG-24-15}

\abstract{
We study two-dimensional $\mathcal{N}{=}(0,2)$ supersymmetric gauged linear sigma models (GLSMs)  using supersymmetric localization. 
We consider $\mathcal{N}{=}(0,2)$ theories with an $R$-symmetry, which can always be defined on curved space by a pseudo-topological twist while preserving one of the two supercharges of flat space. 
For GLSMs which are deformations of $\mathcal{N}{=}(2,2)$ GLSMs and retain a Coulomb branch, we consider the $A/2$-twist and compute the genus-zero correlation functions of certain pseudo-chiral operators, which generalize the simplest twisted chiral ring operators away from the $\mathcal{N}{=}(2,2)$ locus. These correlation functions can be written in terms of a certain residue operation on the Coulomb branch, generalizing the Jeffrey-Kirwan residue prescription relevant for the $\mathcal{N}{=}(2,2)$ locus.
For abelian GLSMs, we reproduce existing results with new formulas that render the quantum sheaf cohomology relations and other properties manifest. For non-abelian GLSMs, our methods lead to new results. As an example, we briefly discuss the quantum sheaf cohomology of the Grassmannian manifold.

}

\begin{document}

\tableofcontents

\section{Introduction}

Supersymmetric localization of the two-dimensional gauged linear sigma model (GLSM) has proven an extremely useful tool in the study of two-dimensional superconformal theories and of string compactifications---see {\it e.g.} \cite{Benini:2012ui,Doroud:2012xw, Jockers:2012dk, Gadde:2013ftv, Benini:2013nda,Benini:2013xpa, Halverson:2013qca,Hori:2013ika} for some of the most important recent progress in that direction.
Most of these recent developments, however, were concerned with theories with $\CN{=}(2,2)$ supersymmetry.~\footnote{The one notable exception is the elliptic genus  computation of \cite{Benini:2013nda,Benini:2013xpa}.}
In the present work, we consider two-dimensional GLSMs with $\CN{=}(0,2)$ supersymmetry defined on $S^2$, assuming that the flat-space theory preserves an $R$-symmetry.
The only way to define a non-conformal supersymmetric $\CN{=}(0,2)$ theory---such as the GLSM---on the sphere  is by a so-called pseudo-topological twist \cite{Witten:1993yc}, which involves a background flux for the $R$-symmetry.~\footnote{
This is to be contrasted with the  $\CN{=}(2,2)$ case, where it is also possible to define a `physical' supersymmetric theory on the sphere without  $R$-symmetry background flux \cite{Benini:2012ui,Doroud:2012xw, Closset:2014pda}.  See also \cite{Bae:2015eoa} for a finer classification of supersymmetric backgrounds on the sphere.}
In the $\CN{=}(2,2)$ case, supersymmetric localization of the $A$-twisted GLSM was recently revisited in \cite{Benini:2015noa,Closset:2015rna}. Here we generalize  these results to the $\CN{=}(0,2)$ world. (See also \cite{McOrist:2007kp, McOrist:2008ji, Melnikov:2009nh} for some previous related work.)

We focus on the case of an $\CN{=}(0,2)$ GLSM with an $\CN{=}(2,2)$ locus---that is, the theory is a continuous deformation of an $\CN{=}(2,2)$ theory, to which it reduces at a special locus in parameter space.
By performing the so-called $A/2$-twist, the theory can be defined on any Riemann surface $\Sigma$ while preserving a single supercharge $\t\CQ_{(A/2)}$. 
Such a theory contains a sector of $\t \CQ_{(A/2)}$-closed operators with non-singular operator product expansions (OPEs)  \cite{Adams:2003zy, Adams:2005tc},  forming what is now known as a `quantum sheaf cohomology' (QSC) ring, generalizing the ordinary quantum cohomology ring (or twisted chiral ring) of the $A$-twisted $\CN{=}(2,2)$ theory.

For $\Sigma \cong \mathbb{P}^1$, correlation functions of 
$\t\CQ_{(A/2)}$-closed operators are topological---they are independent of the insertion points and of the metric on $ \mathbb{P}^1$. The simplest  $\t\CQ_{(A/2)}$-closed operators are the gauge-invariant polynomials  $\CO(\sigma)$ in the $\CN{=}(0,2)$ chiral multiplet scalar $\sigma$, which descents from the scalar field of the $\CN{=}(2,2)$ vector multiplet.
We argue that the correlation functions of these operators
 can be efficiently computed in terms of
`Jeffrey-Kirwan-Grothendieck' (JKG) residues, generalizing the Jeffrey-Kirwan (JK) residue \cite{1993alg.geom..7001J, 1999math......3178B, szenes-vergne}. Schematically, we find
\be\label{JK intro}
\langle \CO(\sigma)\rangle_{\mathbb{P}^1}^{(A/2)} = \sum_k q^k   \oint_{\rm JKG}   \CZ_k^{\rm \oneloop} \, \CO~,
\ee
where the sum is over all the allowed fluxes on the sphere, 
each summand is a particular JKG residue on $\t\fM$, 
the covering space of the GLSM classical Coulomb branch, and $\CZ_k^{\rm \oneloop}$ is a locally holomorphic top form with singularities along divisors on $\t\fM$. The JKG residue is a conjectured residue operation on locally holomorphic forms with prescribed singularities along divisors, and to the best of our knowledge it has not been defined previously in the mathematical literature. We will give our working definition of it in section \ref{subsec: JKG def}. It naturally generalizes the JK residue, which is defined for holomorphic forms with singularities along hyperplanes.
The formula \eqref{JK intro} specializes to the result of \cite{Benini:2015noa, Closset:2015rna}  
for $A$-twisted correlation functions  on the $\CN{=}(2,2)$ locus. We will also briefly discuss a dual version of this formula for some $B/2$-twisted models without an $\CN{=}(2,2)$ locus \cite{Sharpe:2006qd}.

The GLSMs that we consider provide simple ultraviolet (UV) completions of non-linear sigma models (NLSM) on K\"ahler varieties $X$ endowed with an holomorphic vector bundle (more generally, a locally free sheaf) ${\bf E}$ which is a deformation of the tangent bundle $TX$, and reduces to it on the $\CN{=}(2,2)$ locus. The $\t\CQ_{(A/2)}$ cohomology is naturally identified with the sheaf cohomology of ${\bf E}$, and the non-perturbative correlation functions  realize the so-called quantum sheaf cohomology relations.
There has been a considerable amount of previous work on quantum sheaf
cohomology rings in abelian GLSMs, see {\it e.g.} 
\cite{Katz:2004nn,Adams:2003zy,Sharpe:2005fd,Adams:2005tc,Sharpe:2006qd,Guffin:2007mp,McOrist:2007kp,Tan:2006qt,Tan:2007bh,Melnikov:2007xi,Melnikov:2009nh,Kreuzer:2010ph,gs02,Melnikov:2010sa,Aspinwall:2010ve,McOrist:2010ae,Guffin:2011mx,McOrist:2011bn}, 
which culminated in expressions for quantum sheaf cohomology rings for
toric varieties with deformations of the tangent bundle, derived both
from physics in \cite{McOrist:2008ji} and from mathematics in 
\cite{Donagi:2011uz,Donagi:2011va}.  (See for example
\cite{Melnikov:2012hk,Garavuso:2013zoa,Sharpe:2015vza,Lu:thesis,Lu:2015a} 
for more recent developments and reviews.)  
In abelian examples, part of the appeal
of our methods is that it gives more efficient computational methods
for correlation functions than have existed previously.
Another appeal of localization is that it makes manifest some previously
obscure properties of correlation functions, namely their independence from nonlinear
deformations, and the independence of $A/2$-twisted correlation functions from
$J$-type bundle deformations. (Analogously, the $B/2$-twisted correlation functions that we will consider are independent of $E$-type deformations.)

Furthermore, our methods also extend to non-abelian GLSMs, which were
intractable with previous methods.   As an example, we consider the $\CN{=}(0,2)$ GLSM for the Grassmannian manifold with a deformed tangent bundle and compute the  $A/2$-twisted correlation functions. This leads to a prediction for the quantum sheaf cohomology of this model,  which will be studied further in  \cite{Guo:2015caf, gls2}.

The formula \eqref{JK intro}  passes some strong consistency checks.
In the abelian cases, it encodes explicitly the
quantum sheaf cohomology relations,  in agreement with previous results
\cite{McOrist:2008ji,Donagi:2011uz,Donagi:2011va}.
In addition, we can compare the results     obtained from  \eqref{JK intro}  for the simplest correlation functions of the  ${\mathbb P}^1 \times {\mathbb P}^1$ and 
${\mathbb F}_1$ models to the NLSM expressions obtained using older \v{C}ech-cohomology-based
methods, and we find perfect agreement.

This paper is organized as follows. In section \ref{sec: curved space susy}, we study curved-space supersymmetry for $\CN{=}(0,2)$ theories with an $R$-symmetry, and we discuss GLSMs in particular.  In section \ref{sec: loc in GLSM 1}, we specialize to $A/2$-twisted GLSMs with an $\CN{=}(2,2)$ locus,  we derive our main result \eqref{JK intro} and we discuss a few of its  consequences. In section \ref{sec: Abelian expl}, we apply the JKG residue formula to some well-studied abelian models. In section \ref{sect:nonabelian}, we consider the simplest examples of non-abelian GLSMs, namely the Grassmannian manifold and complete interesection Calabi-Yau manifolds inside the Grassmannian. In section \ref{sec: generalizations}, we briefly discuss a generalization of our main formula to theories with `twisted masses' and to $B/2$-twisted GLSMs dual to the $A/2$-twisted models of section  \ref{sec: loc in GLSM 1}. Some useful auxiliary material can be found in appendices.

\section{$\CN{=}(0,2)$ curved-space supersymmetry}\label{sec: curved space susy}
We wish to consider  $\CN{=}(0,2)$ supersymmetric gauge theories with an $R$-symmetry, denoted $U(1)_R$. In this section, we explain how to preserve supersymmetry on any closed orientable Riemann surface $\Sigma_{\bf g}$.  We then discuss  $\CN{=}(0,2)$ supersymmetric multiplets, Lagrangians and observables on curved space.  We refer to appendix \ref{app: conventions} for a summary of our curved-space conventions, and for a review of $\CN{=}(0,2)$ supersymmetry in flat space.

\subsection{Background supergravity and the pseudo-topological twist} 
Consider any $\CN{=}(0,2)$ supersymmetric field theory with an $R$-symmetry. The theory possesses  a conserved $\CR$-symmetry current $j_\mu^{(R)}$ which sits in the $\CN{=}(0,2)$ $\CR$-multiplet  \cite{Dumitrescu:2011iu} together with the right-moving supercurrent $S^\mu_+$, $\t S^\mu_+$ and the energy-momentum tensor $T_{\mu\nu}$.  Such a theory can be coupled to an $\CN{=}(0,2)$ background supergravity multiplet containing a metric $g_{\mu\nu}$, two gravitini $\psi_{-\mu}$, $\t\psi_{-\mu}$ and a $U(1)_R$ gauge field $A_\mu^{(R)}$. At first order around flat space, $g_{\mu\nu}= \delta_{\mu\nu}+ \Delta g_{\mu\nu}$, the supergravity multiplet couples to the $\CR$-multiplet according to:
\be\label{Lag sugra ii}
\SL_{SUGRA}=  - \half \Delta g_{\mu\nu} T^{\mu\nu}+ A_\mu^{(R)}  j^\mu_{(R)} -\half \left(S^\mu_+ \psi_{-\mu}-\t S^\mu_+ \t\psi_{-\mu}   \right)~. 
\ee
Curved-space rigid supersymmetry is best understood in terms of a supersymmetric background for the metric and its superpartners \cite{Festuccia:2011ws, Adams:2011vw}. A background $(\Sigma_{\bf g}, g_{\mu\nu}, A_\mu^{(R)})$ is supersymmetric if and only if  the supersymmetry variations of the gravitini vanish for some non-trivial supersymmetry parameters. In the present case,  we must have:
\be\label{KSE 02}
(\nabla_{\mu}  - i A_{\mu}^{(R)})\zeta_ -=0~, \qquad \qquad (\nabla_{\mu}  + i A_{\mu}^{(R)})\t\zeta_ -=0~.
\ee
Note that the spinors $\zeta_-$, $\t\zeta_-$ have $R$-charge $\pm 1$, respectively.  One can derive these equations by studying linearized supergravity along the lines of \cite{Closset:2014pda}. (See also \cite{Gomis:2015yaa} for a complementary discussion.)
The only way to solve \eqref{KSE 02}  on  $\Sigma_{\bf g}$  is by setting the gauge field $A_\mu=\mp \half \omega_\mu$, with $\omega_\mu$ the spin connection. This  preserves either $\zeta_-$ or $\t\zeta_-$.  (The only obvious exception is when $\Sigma_{{\bf g}=1}$ is a flat torus.)  We  choose to preserve $\t\zeta_-$:
\be\label{half twist background}
A^{(R)}_\mu = \half \omega_\mu~, \qquad \quad \zeta_-=0~, \qquad \d_\mu\t\zeta_- =0~.
\ee
Since $\t\zeta_-$ is a constant,  it is obviously well-defined globally on $\Sigma_{\bf g}$. 
This supersymmetric background corresponds to a pseudo-topological twist \cite{Witten:1993yc} and it preserves one supercharge $\t Q_+$ on any $\Sigma_{\bf g}$. 
It follows from \eqref{half twist background} that
\be
{1\over 2\pi}\int_{\Sigma} dA^{(R)} =- {1\over 8\pi}\int_{\Sigma} d^2 x\sqrt{g} \, {\rm R}= g-1~,
\ee
where  ${\rm R}$ is the Ricci scalar  of $g_{\mu\nu}$, and therefore  the $R$-charge is quantized in units of ${1\over g-1}$. In particular,  the $R$-charge is integer-quantized on the Riemann sphere.

\subsection{Supersymmetry multiplets} 
Since the supersymmetry parameter $\t\zeta_-$ is covariantly conserved, the supersymmetry variations in curved space can be  obtained from the flat space expressions by replacing derivatives by covariant derivatives. Let us denote by $\delta$ the supercharge $\t Q_+$ acting on fields. Importantly, $\delta$ is nilpotent:
\be
\delta^2=0~.
\ee
The pseudo-topological twist effectively assigns to every field a spin
\be\label{top twist}
S= S_0+\half R~,
\ee
where $S_0$ and $R$ are the flat-space spin and the flat-space $R$-charge, respectively.
The twist \eqref{top twist} can correspond to any of the distinct twists that one might define in a given theory, corresponding to distinct choices for the $R$-symmetry. 

 It is convenient to use a notation adapted to the twist, in terms of which all the fields have vanishing $R$-charge and definite twisted spin. 
 We use the covariant derivatives
\be
D_\mu \varphi_{(s)} = \left(\d_\mu -i{s}\omega_\mu\right)\varphi_{(s)}~,
\ee
acting on a field of  twisted spin $s$.  We summarize   our curved-space conventions, as well as the relation between flat-space and twisted variables,  in appendix \ref{app: conventions}.

\subsubsection{General multiplet}
Let $\CS_s$ be a general   multiplet of $\CN{=}(0,2)$ supersymmetry with $2+2$ complex components, with $s$ the twisted spin of the lowest component:
\be\label{gen multiplet Sigma}
\CS_{s} = \left({\bf C}~,\, \chi_{\b1}~,\, \t\chi~, \, {\bf v}_{\b1}\right)~.
\ee
The four components of \eqref{gen multiplet Sigma} have spin $(s, s-1, s, s-1)$, respectively. The curved-space supersymmetry transformations are:
\bea\label{susy gen}
&\delta {\bf C} = - i \t\chi~, \qquad\quad & & \delta \chi_{\b1} = 2 i {\bf v}_{\b1} + 2 D_{\b1} {\bf C}~,\cr
&\delta \t\chi=0~, \qquad && \delta {\bf v}_{\b1}= D_{\b1} \t\chi~.
\eea
Note that $\delta$ is a scalar---it commutes with the spin operator.  All the supersymmetry multiplets of interest to us are made out of one or two general multiplets subject to some conditions. 

\subsubsection{Chiral multiplets}
The simplest $\CN{=}(0,2)$ multiplets are the chiral multiplet $\Phi_i$ and the antichiral multiplet $\t\Phi_i$.  In flat space, they contains a complex scalar and a spin $-\half$ fermion. After twisting, one has:
\be\label{phi tphi components}
\Phi_i= \left(\phi_i~,\,  \CC_{  i}\right)~, \qquad \qquad \t\Phi_i= \left(\t\phi_i~,\,  \t\CB_i\right)~.
\ee
If $\Phi_i$, $\t\Phi_i$ are assigned integer $R$-charges $r_i$ and $-r_i$, the components \eqref{phi tphi components} have twisted spins $({r_i\over 2}, {r_i\over 2}-1)$ and $(-{r_i\over 2},-{r_i\over 2})$, respectively. The supersymmetry transformations rules are:
\bea\label{susy twisted chiral}
&\delta \phi_i = 0~,\qquad \qquad \qquad   \qquad && \delta \t\phi_i = \t\CB_i~, \cr
&  \delta \CC_{ i}  = 2 i D_{\b1} \phi^i ~,\qquad   \qquad && \delta  \t\CB_i =0~.
\eea
Note that $\Phi$ and $\t\Phi$ can be understood as a general multiplets \eqref{gen multiplet Sigma} satisfying the constraints $\t\chi=0$ and $\chi_{\b 1}=0$, respectively. Here and in the following, the $R$-charge refers to the flat-space $R$-charge, since the twisted variables used in curved space have vanishing $R$-charge by construction.

Given any holomorphic function $\CF(\Phi_i)$ of the chiral multiplets $\Phi_i$, one can construct a new chiral multiplet as long as $\CF$ itself has definite $R$-charge, and  similarly with the anti-chiral multiplets:
\be
 (\CF~,\,   \CC^{\CF})= \left(\CF(\phi)~,\, {\d \CF\ov \d \phi_i} \CC_i\right)~, \qquad\quad
(\t \CF,  \t\CB^\CF)= \left(\t \CF(\t\phi)~,\, {\d \t \CF\ov \d \t\phi_i} \t\CB_i\right)~.
\ee

\subsubsection{Fermi multiplets}
Another important multiplet is the Fermi multiplet, whose lowest flat-space component is a spin $+\half$ fermion. For each Fermi multiplet $\Lambda_{I}$, we have a function $\CE_I(\phi)$ holomorphic in the chiral fields of the theory. Similarly, an anti-Fermi multiplet $\t\Lambda_I$ comes with an anti-holomorphic function $\t\CE(\t\phi)$. For the elementary Fermi multiplets,  these functions must be specified as part of the data defining the $\CN{=}(0,2)$ theory. In order to preserve the $R$-symmetry, they must have $R$-charges $R[\CE_I]= r_I+1$, with $r_I$ is the $R$-charge of $\Lambda$. Similarly, the charge-conjugate multiplet $\t\Lambda$ has $R$-charge $-r_I$ and $R[\t\CE_I]= -r_I-1$.

A Fermi multiplet $\Lambda_I$ of $R$-charge $r_I$ has components:
\be\label{fermi twisted compo i}
\Lambda_I =\left(\Lambda_I~,\,  \CG_I\right)~, \qquad E_I= \left(\CE_I, \CC^E_{ I}\right)~,
\ee
where $E_I$ is the chiral multiplet of lowest component $\CE_I$. The spins of \eqref{fermi twisted compo i} are $({r_I\ov 2}+\half, {r_I\ov 2}-\half)$ and $({r_I\ov 2}+\half, {r_I\ov 2}-\half)$, respectively, and the supersymmetry transformations are given by:
\bea\label{fermi twisted var}
&\delta \Lambda_I = 2 \CE_I~, \qquad \qquad &&\delta \CE_I =0~, \cr
&\delta \CG_I= 2 \CC_I^E - 2 i D_{\b1} \Lambda_I~, \qquad \qquad &&\delta \CC_I^E = 2 i D_{\b1}\CE_I~.
\eea 
Similarly, for an anti-Fermi multiplet $\t\Lambda_I$ of $R$-charge $- r_I$, we have the components:
\be\label{fermi twisted compo ii}
\t\Lambda_I = \left(\t\Lambda_I~,\, \t\CG_I\right)~, \qquad \t E_I= \left(\t\CE_I~,\, \t \CB^E_I\right)~,
\ee
of spins $(-{r_I\ov 2}+\half, -{r_I\ov 2}+\half)$ and $(-{r_I\ov 2}-\half, -{r_I\ov 2}-\half)$, respectively, and
\bea\label{antifermi twisted var}
&\delta\t \Lambda_I = \t\CG_I~, \qquad \qquad &&\delta \t\CE_I =\t\CB_I^E~, \cr
&\delta \t\CG_I=0~, \qquad \qquad &&\delta \t\CB_I^E =0~.
\eea
The product of a chiral multiplet $\Phi_i$ of $R$-charge $r_i$ with a Fermi multiplet $\Lambda_I$ of $R$-charge $r_I$ gives another Fermi multiplet of $R$-charge $r_i+ r_I$, with components:
\be
\Lambda^{(\Phi \Lambda)} =\left(\phi_i\Lambda_I~,\,  \phi_i\CG_I- \CC_i \Lambda_I\right)~, \qquad E_I^{(\Phi\Lambda)}= \left(\phi_i \CE_I~, \phi_i \CC^E_{ I} + \CC_i \CE_I \right)~.
\ee
Similarly, for the charge-conjugate multiplet:
\be
\t\Lambda^{(\Phi \Lambda)}  = \left(\t\phi_i \t\Lambda_I~,\, \t\phi_i \t\CG_I + \t\CB_i \t\Lambda_I \right)~, \qquad \t E_I^{(\Phi\Lambda)}= \left(\t\phi_i \t\CE_I~,\, \t\phi_i \t \CB^E_I + \t\CB_i \t\CE_I^E\right)~.
\ee

\subsubsection{Vector multiplet}
Consider a compact Lie group $\GG$ and its Lie algebra $\Fg$. 
The associated vector multiplet is built out of two $\Fg$-valued general multiplets $(\CV, \CV_{1})$ of spins $0$ and $1$, subject to the gauge transformations:
\be\label{deltaOmega V}
\delta_\Omega \CV ={i\ov 2} (\Omega-\t\Omega) + {i\ov 2}[\Omega+\t\Omega, \CV]~, \qquad\quad
\delta_\Omega \CV_{1} =\half \d_1(\Omega+\t\Omega) + {i\ov 2}[\Omega+\t\Omega, \CV_1]~,
\ee
where $\Omega$ and  $\t\Omega$ are $\Fg$-valued chiral and  antichiral multiplets of vanishing $R$-charge.~\footnote{The addition rules implicit in \eqref{deltaOmega V} are obtained by embedding $\Omega$, $\t\Omega$ into general multiplets.}
One can use \eqref{deltaOmega V} to fix a Wess-Zumino (WZ) gauge, wherein  the vector multiplet has  components:
\be\label{vec mult02}
\CV = \left(0~,\, 0~,\, 0~, \, a_{\b1}\right)~, \qquad \CV_{1}= \left(a_{1}~,\,   \t\lambda~,\, \lambda_1, \, D\right)~,
\ee
The non-zero components have spin $-1$ and $(1, 0, 1, 0)$, respectively. Under the residual gauge transformations $\Omega=\t\Omega=(\omega, 0)$, we have
\be
\delta_\omega a_\mu = \d_\mu \omega + i[\omega, a_\mu]~, \quad
\delta_\omega \lambda_1 = i [\omega, \lambda_1]~, \quad
\delta_\omega \t\lambda = i [\omega, \t\lambda]~, \quad
\delta_\omega D = i [\omega, D]~.
\ee
 The supersymmetry transformations are:
\bea
& \delta a_{\b1}=0~, \qquad &&\delta a_1 = - i \lambda_1~, \qquad & \cr 
 & \delta\t\lambda= - i (D- 2 i f_{1\b1})~, \qquad\quad && \delta\lambda_1=0~, \qquad\quad\quad
&\delta D= - 2 D_{\b1}\lambda_1~,
\eea
where we defined the field strength
\be\label{def f11b}
f_{1\b 1}= \d_1a_{\b1} - \d_{\b1} a_1 - i[a_1, a_{\b1}] ~,
\ee
and the covariant derivative $D_\mu$  is also gauge-covariant.
Here and henceforth, $\delta$ denotes the supersymmetry variation in WZ gauge, which includes a compensating gauge transformation.

\subsubsection{Field strength multiplet}
From the vector multiplet \eqref{vec mult02}, one can build a Fermi and an anti-Fermi multiplet:
\be\label{def Y tY}
\CY= \Big(2 \lambda_1~, \, 2 i (2 i f_{1\b 1})\Big)~, \qquad \qquad \t\CY= \Big(\t \lambda~, \, - i (D-2 i f_{1\b 1})\Big)~,
\ee
of $R$-charge $1$ (that is, the multiplets $\CY$ and  $\t\CY$ have twisted spin $1$ and $0$, respectively), with $\CE_\CY=0$. These field strength multiplets are $\Fg$-valued.~\footnote{Our definition of $\CY$ in \eqref{def Y tY}  a slightly idiosyncratic. There is a unique definition for $\CY$ in flat space, namely $(2 \lambda_1~, \, i (D+2 i f_{1\b 1}))$, but in curved space with one supercharge the present choice is also consistent. The present choice is the same as in \cite{Closset:2015rna}. \label{footnote Y}} 

\subsubsection{Charged chiral and Fermi multiplets}
Consider the chiral multiplets $\Phi_i$ in the representations $\FR_i$ of the gauge algebra $\Fg$, the Fermi multiplets $\Lambda_I$ in the representations $\FR_I$ of $\Fg$, and similarly for the charge conjugate multiplets $\t\Phi_i$ and $\t\Lambda_I$.  Under a gauge transformation \eqref{deltaOmega V}, we have
\be
\delta_\Omega \Phi=  i \Omega \Phi~, \qquad \delta_\Omega \t\Phi= - i \t \Phi \t\Omega~, \qquad \quad  \delta_\Omega \Lambda=  i \Omega \Lambda~, \qquad   \delta_\Omega \t\Lambda=  -i\t \Lambda\t \Omega~,
\ee
with $\Omega$, $\t\Omega$ valued in the corresponding representations.  The supersymmetry transformations in WZ gauge are given by \eqref{susy twisted chiral}, \eqref{fermi twisted var} and \eqref{antifermi twisted var} with the understanding that the covariant derivative $D_\mu$ is   also gauge-covariant.

\subsubsection{Conserved current and background vector multiplet}
Consider a theory with a global continuous symmetry group $\GG^F$.  The corresponding conserved current ${\bf j}^\mu$ sits in multiplet
\be\label{twisted conserved current CJ}
\CJ=  \left(J~,\, j_\bz~,\, \t j~,\, {\bf j}_z~,\,  {\bf j}_\bz\right)
\ee
which is built out of two general multiplets of spin $0$ and $1$. The components \eqref{twisted conserved current CJ} have twisted spins $(0,-1,0,1,-1)$, respectively. $J$ is a bosonic scalar operator, $j_\bz$ and $\t j$  are fermionic, and ${\bf j}_\mu$ satisfies
\be
D_z {\bf j}_\bz+ D_\bz {\bf j}_z = 0~.
\ee
The supersymmetry transformations are 
\bea
&\delta J= -i \, \t j~, \qquad \delta j_\bz = 2 (\d_\bz J - i {\bf j}_\bz)~, \qquad \delta \t j=0~, \cr
&\delta {\bf j}_z = \d_z \t j~, \qquad \;\delta {\bf j}_\bz =- \d_\bz \t j~.
\eea
Such a conserved current can be coupled to a (background) vector multiplet. At first order in the gauge field, we have:
\be
\SL_{\CV \CJ}  = a_\mu {\bf j}^\mu + D J + ({\rm fermions})~.
\ee

\subsection{Supersymmetric Lagrangians}
There are four types of supersymmetric Lagrangians we can consider on curved space:

\begin{itemize}
\item[1.] {\it ${\bf v}$-term.}  Given a general multiplet $\CS_{1}$ of twisted spin  $s=1$ with components \eqref{gen multiplet Sigma}, we can built the supersymmetric Lagrangian
\be
\SL_{\bf v}= {\bf v}_{1\b1}
\ee
from the top component. It is clear from \eqref{susy gen} that the corresponding action is both $\delta$-closed and $\delta$-exact.

\item[2.] {\it $\CG$-term.} From a Fermi multiplet $\Lambda$ with $s=1$ (that is, $R$-charge $\half$) and $\CE=0$, we have the supersymmetric Lagrangian
\be\label{G term Lag} 
\SL_{\CG}= \CG~.
\ee
This term is {\it not} $\delta$-exact.

\item[3.] {\it $\t\CG$-term.}  From an anti-Fermi multiplet $\t\Lambda$ with $s=0$ (that is, $R$-charge $\half$), we can similarly build
\be\label{tG term Lag}
\SL_{\t\CG}= \t\CG~.
\ee
We  see from \eqref{antifermi twisted var} that this term is both $\delta$-closed and $\delta$-exact.  

\item[4.]  {\it Improvement Lagrangian.} This term is special to curved space. Given a conserved current multiplet \eqref{twisted conserved current CJ}, the Lagrangian
\be\label{impr lag}
\SL_{\CJ} = A^{(R)}_\mu {\bf j}^\mu +{1\over 4} {\rm R} J~,
\ee
 is supersymmetric upon using \eqref{half twist background}. .
\end{itemize}
In the remainder of this section, we spell out the various Lagrangians that we shall need later on.

\subsubsection{Kinetic terms}\label{subsec: kin}

All the standard kinetic terms are ${\bf v}$-terms and are  therefore $\delta$-exact. Consider a $\Fg$-valued vector multiplet. The standard supersymmetric Yang-Mills Lagrangian reads:
\be\label{SYM lag}
\SL_{YM} = {1\over e_0^2}\left( \half (2i f_{1\b1})^2 -\half D^2 - 2i \t\lambda D_{\b1} \lambda_1 \right)~.
\ee
Here and below, the appropriate trace over $\Fg$ is implicit. The Lagrangian \eqref{SYM lag} is  $\delta$-exact:
\be
\SL_{YM} =  {1\over e_0^2} \delta\left( {1\ov 2 i} \t\lambda (D+ 2 i f_{1\b1}) \right)~.
\ee
Consider   charged chiral multiplets $\Phi_i$ of $R$-charges $r_i$, transforming in representation $\FR_i$ of $\Fg$. Their kinetic term reads
\be\label{Lag phi i}
\SL_{\t\Phi\Phi} = D_\mu\t\phi^i D^\mu\phi_i +{r_i\ov 4}{\rm R} \t\phi^i \phi_i + \t\phi^i D\phi_i + 2 i \t\CB^i D_1 \CC_i - 2i \t\phi^i\lambda_1\CC_i + i\t\CB^i\t\lambda \phi_i~,
\ee
 where the vector multiplet fields $(a_\mu,  \t\lambda, \lambda_1, D)$ are suitably   $\FR_i$-valued. The Lagrangian \eqref{Lag phi i} is more conveniently written as:
\bea
&\SL_{\t\Phi\Phi} &=&\;  \delta\left(2 i  \t\phi^i D_1\CC_i + i \t\phi^i \t\lambda\phi_i \right)\cr
&&=&\; \t\phi^i\left(- 4 D_1 D_{\b1} + D - 2if_{1\b1} \right)\phi_i+ 2 i \t\CB^i D_1 \CC_i - 2i \t\phi^i\lambda_1\CC_i + i\t\CB^i\t\lambda \phi_i~.
\eea
Similarly, for  charged Fermi multiplets $\Lambda_I$ of $R$-charges $r_I$ in  representations $\FR_I$ of $\Fg$, we have
\bea\label{Fermi kin lag}
&\SL_{\t\Lambda\Lambda} &=& \;  \delta\left( -\t\Lambda^I \CG_I +\half \t\CE^I\Lambda_I \right)\cr
&&=&\; - 2i \t\Lambda^I D_{\b1}\Lambda_I - \t\CG^I\CG_I + \t\CE^I\CE_I + 2\t\Lambda^I {\d \CE_I\ov \d \phi^i} \CC^i +\half \t\CB^i {\d\t\CE^I\ov \d \t\phi^i} \Lambda_I~,
\eea
including the $\FR_I$-valued gauge field in the covariant derivatives $D_{\b 1}$. The holomorphic functions $\CE_I(\phi)$  transform in the same  representations $\FR_I$ as $\Lambda_I$.

\subsubsection{Superpotential terms}
To each Fermi multiplet $\Lambda_I$, one can associate a holomorphic function of the chiral multiplets  $J_I=J_I(\Phi)$, transforming in the representation $\b\FR_I$ conjugate to $\FR_I$ and with $R$-charge $1-r_I$. From these $\CN{=}(0,2)$ superpotential (or $J$-potentials), one can build  the  $\CG$-term Lagrangian \eqref{G term Lag} according to:
\be\label{J term lag}
\SL_{J_I} = i \sum_I \CG^{(J_I)} =i \CG^I J_{I}  + i \Lambda^I {\d J_I \ov \d \phi^i} \CC^i~.
\ee
Note that this Lagrangian is not $\delta$-exact. Supersymmetry implies that
\be\label{JE constraint}
\CE^I J_I =0~.
\ee
Similarly, from the charge conjugate anti-holomorphic functions $\t J_I=\t J_I(\t\Phi)$ one builds the $\t \CG$-term:
\be
\SL_{\t J_I} = -i \sum_I  \t\CG^{(\t J_I)} =-i \t\CG^I \t J_{I}  + i \t \Lambda^I {\d \t J_I \ov \d \t \phi^i} \t\CB^i  = \delta\left( - i \t\Lambda^I \t J_I\right)~,
\ee
which is $\delta$-closed and $\delta$-exact.

\subsubsection{Fayet-Iliopoulos terms}
Consider a gauge theory with Abelian factors $U(1)_A \subset \GG$. From \eqref{def Y tY}, we construct the gauge invariant Fermi multiplets
\be
\CY_A = \tr_A(\CY)~, \qquad \t\CY_A=\tr_A(\t\CY)~,
\ee
where $\tr_A$ is the projection onto the $U(1)_A$ factor. These Fermi multiplet have vanishing $\CE$-potential but they admit $J$-potentials. In the present work, we restrict ourselves to the case of a constant $J_{\CY_A}=J_A$ in the classical Lagrangian:
\be\label{FI JA terms}
J_A = \tau_A \equiv {\theta_A\over 2\pi} + i\xi_A~, \qquad \qquad
\t J_A =\t \tau_A \equiv - 2 i\xi_A~.
\ee
Here $\xi_A$ and $\theta_A$ are the Fayet-Iliopoulos (FI) and $\theta$-angles, respectively. (The unusual definition of $\t\tau$ is on par with \eqref{def Y tY}.) The corresponding supersymmetric Lagrangian reads
\be
\CL_{FI} = \half \left(\tau \CG^\CY+ \t\tau \t\CG^{\t\tau}\right) =  i {\theta^A\ov 2\pi}\tr_A(2i f_{1\b 1}) -\xi^A \tr_A(D)~.
\ee
Note that the coupling $\t\tau$  is $\delta$-exact while the coupling $\tau$  is not.

\subsubsection{Supersymmetric counterterm}
We can build a trivially-conserved current multiplet \eqref{twisted conserved current CJ} from 
\be
J= f(\phi) + \t f(\t \phi)~,
\ee
with $f(\phi)$ and $f(\t\phi)$ some (anti)holomorphic functions of the (anti)chiral multiplet lowest components, of vanishing $R$-charge. The improvement Lagrangian  \eqref{impr lag} reads:
\be\label{local ctterm}
\SL_{\rm ct} = {1\over 2} {\rm R} f(\phi)
\ee
in this case.  Note that the dependence on the anti-holomorphic function $\t f$ dropped out. The Lagrangian \eqref{local ctterm} is therefore an purely holomorphic local term on the twisted sphere. 


\subsection{GLSM field content and anomalies}

Consider a general $\CN{=}(0,2)$ GLSM with a gauge group $\GG$, with $\Fg= {\rm Lie}(\GG)$.
  The gauge sector consists of a  $\Fg$-valued vector multiplet $(\CV, \CV_1)$. If $\GG$ contains $U(1)$ factors,
\be\label{U1A factors}
\prod_{A=1}^n U(1)_A \subset \GG~,
\ee
we turn on the FI parameters \eqref{FI JA terms}. Let us also define the quantity:
\be\label{def qA}
q_A = \exp(2\pi i \tau_A)~.
\ee
The matter sector consists of chiral multiplets $\Phi_i$ of $R$-charges $r_i$ in representations  $\FR_i$ of $\Fg$, and of  Fermi multiplets $\Lambda_I$ of $R$-charges $r_I$ in representations $\FR_I$ of $\Fg$. 
To each $\Lambda_I$, we associate the two holomorphic potentials $\CE_I= \CE_I(\Phi)$ and $J_I = J_I(\Phi)$ constructed out of  the chiral multiplets $\Phi_i$, satisfying $\CE^I  J_I= 0$,
 with $R$-charges
\be
R[\CE_I]= r_I +1~, \qquad R[J_I]= 1 -r_I~,
\ee
and such that $\Tr(\t\Lambda^I \CE_I)$ and $\Tr(\Lambda^I J_I)$ are gauge invariant.

Anomaly cancelation imposes  further constraints on the matter content and on the $R$-charge assignment. Let us decompose the gauge algebra $\Fg$ into semi-simple factors $\Fg_\alpha$ and Abelian factors $u(1)_A$, $\Fg \cong (\oplus_\alpha\Fg_\alpha) \oplus (\oplus_A u(1)_A)$. The vanishing of the non-Abelian gauge anomalies requires 
\be
\sum_i  T_{\FR_i^{(\alpha)} }  -\sum_I  T_{\FR_I^{(\alpha)} } - T_{\Fg_{\alpha} } =0~, \qquad \forall \alpha~,
\ee
where $\FR^{(\alpha)} $ denotes the  representation of $\Fg_\alpha$ obtained by projecting the representation $\FR$ of $\Fg$ onto $\Fg_\alpha$, while $T_{\FR^{(\alpha)} }$ denotes the Dynkin index of $\FR^{(\alpha)}$ and $T_{\Fg_{\alpha} } $ stands for the index of the adjoint representation of $\Fg_\alpha$. For instance, one has $T_{\rm fund }= T_{\overline{\rm fund }}=\half$ and  $T_{su(N)}=N$ for the fundamental, antifundamental  and adjoint representations of $su(N)$.  In order to cancel the $U(1)^2$ gauge anomalies, we also need
\be
\sum_i \dim{\FR_i}  \, Q_i^A \,  Q_i^B  -\sum_I  \dim{\FR_I} \,  Q_I^A \,  Q_I^B=0~, \qquad \forall A, B~,
\ee
where $Q^A_i$ and $Q_I^A$ are the $U(1)_A$ charges of the chiral and Fermi multiplets, respectively.
In addition,  the $U(1)_R$-gauge anomalies  should vanish:
\be\label{R nonanomalous}
\sum_i \dim{\FR_i} \, (r_i -1) \,Q_i^A  - \sum_I \dim{\FR_I}\,  r_I \, Q_I^A =0~, \qquad \forall A~.
\ee
Let us also note that the FI parameters $\xi_A$ often run at one-loop with $\beta$-functions:
\be\label{beta function xi}
\beta^A \equiv \mu {d\tau^A\over d\mu} = -{b_0^A\over 2\pi i}~,\quad \qquad \quad b_0^A = \sum_i \tr_{\mathfrak{R}_i} (t_A)~,
\ee
 due to contributions from the charged chiral multiplets.

\subsection{Pseudo-topological observables}
Consider an $\CN{=}(0,2)$ theory in curved space, with a certain twist by the $R$-symmetry. The flat-space theory has an $\CR$-multiplet \cite{Dumitrescu:2011iu} that includes the  stress-energy tensor $T_{\mu\nu}$ and the $R$-symmetry current $j_\mu^{(R)}$. We can define a ``twisted'' stress-energy tensor:
\be
\CT_{z z} = T_{zz} - {i\over 2} \d_z j_z^{(R)}~, \qquad
\CT_{z \bz} = T_{z\bz} - {i\over 2} \d_z j_\bz^{(R)}~, \qquad
\CT_{\bz\bz}=T_{\bz\bz} + {i\over 2} \d_\bz j_\bz^{(R)}~,
\ee
which is conserved because $T_{\mu\nu}$ and $j_\mu^{(R)}$ are conserved. 
The operator $\CT_{zz}$ is $\t Q_+$-closed, while $\CT_{z \bz}$ and $\CT_{\bz \bz}$ are also $\t Q_+$-exact. By a standard arguments, it follows that correlation functions of $\t Q_+$-closed operators are independent of the Hermitian structure on the two-dimensional manifold $\Sigma_{\bf g}$, while they may depend holomorphicaly on its complex structure moduli \cite{Witten:1993yc}.

The supersymmetric observables are also (locally) holomorphic functions of the various couplings. It is clear that they are holomorphic in the superpotential couplings appearing in $J_I$, and in the FI parameters $J_A=\tau_A$, since the anti-holomorphic couplings $\t J_I$ and $\t J_A$ are $\delta$-exact. 
To understand the dependence on the $\CE_I$-potential couplings, note that any deformation of $\t\CE_I$ by $\Delta\t\CE_I(\t\phi)$
deforms the classical Lagrangian \eqref{Fermi kin lag} by a $\delta$-exact operator:
\be
\Delta \SL= \Delta\t\CE^I \CE_I+\half \CB^i \d_i(\Delta\t\CE_I) \, \Lambda^I 
= \half  \delta\left(\Delta\t\CE^I \Lambda_I \right)~.
\ee
More generally, it follows from \eqref{antifermi twisted var} that $\t\CE$-deformations commute with the supersymmetry. On the other hands, deformations of the holomorphic potentials $\CE_I$ commute with the supercharge up to terms holomorphic in $\Delta\CE_I$. Since $\CE_I$ only enters the Lagrangian through $\delta$-exact terms, this implies that supersymmetric observables depend holomorphically on the $\CE_I$-couplings. (See \cite{Closset:2014uda} for a similar discussion in four dimensions.)

We are interested in a special class of $\t Q_+$-closed operators with non-singular OPEs \cite{Adams:2003zy, Adams:2005tc}, and we would like to consider  their correlations functions on the Riemann sphere:
\be\label{corr gen twist}
\langle  \CO_a \CO_b \cdots \rangle_{\mathbb{P}^1}~.
\ee
These `pseudo-chiral' operators form a ring with product structure
\be\label{product QSC}
\CO_a\, \CO_b = {f_{ab}}^c\, \CO_c
\ee
captured by the  genus-zero correlation functions. When the GLSM flows at intermediate energies to a NLSM with target space $X$  endowed with an holomorphic vector bundle ${\bf E}$ (more generally, a locally free sheaf), the operators $\CO_a$ are expected to flow to  NLSM operators corresponding to sheaf cohomology classes of  the bundle ${\bf E}$.  In that case, the correlation functions \eqref{corr gen twist} define a quantum-deformed sheaf cohomology ring, known as  quantum sheaf cohomology (QSC).~\footnote{This expectation is satisfied in many simple examples, but the situation can be more complicated. In general, many of the quantum sheaf cohomology classes of ${\bf E}$ are not realized in any simple (or known) way in the GLSM language.} The  QSC relations can be computed in the GLSM in the UV because the pseudo-topological correlators \eqref{corr gen twist} are RG-invariant. (See also \cite{Dedushenko:2015opz} for a recent discussion.)  By abuse of notation, we  sometimes use the term `quantum sheaf cohomology' for the pseudo-chiral ring of a GLSM, irrespective of its geometric interpretation.

In the next section, we will further restrict ourselves to the case of the $A/2$-twisted pseudo-chiral ring of $\CN{=}(0,2)$ theories with an $\CN{=}(2,2)$ locus, while some simple $B/2$-twisted theories will be considered in section \ref{sec: generalizations}. We leave more general studies of arbitrary $\CN{=}(0,2)$ pseudo-chiral rings for future work.

\subsection{Supersymmetric locus and zero-modes on the sphere}
A configuration of bosonic fields from the vector, chiral and Fermi multiplets preserves the single supercharge on curved space if and only if the fields satisfy the supersymmetry equations:
\be\label{susy equations i}
D= 2 i f_{1\b 1}~, \qquad \quad D_{\b z} \phi_i=0~, \qquad \quad \CE_I(\phi) =\t\CE_I(\t\phi) = 0~. 
\ee
In particular, the chiral field $\phi_i$ is an holomorphic section of an holomorphic vector bundle determined by its $R$- and gauge-charges. Such configurations will dominate the path integral. In the special case of an $A/2$-twisted GLSM with an $\CN{=}(2,2)$ locus---to be discussed in the next section---we will argue that  the path integral for pseudo-topological supersymmetric observables can be further localized  into  Coulomb branch configurations, in which case the charged chiral multiplets are massive and localize to $\phi_i=0$. We still have to sum over all the topological sectors, with fluxes:
\be\label{flux sector}
 {1\ov 2 \pi} \int d^2 x \sqrt{g}\, (- 2 i f_{1\b 1})\equiv k \in i \Fh~. 
\ee

Note that we generally have fermionic zero modes, in addition to the bosonic zero modes that solve the second equation in \eqref{susy equations i}.
For future reference, let us summarize the counting of zero modes on the Riemann sphere. (The generalization to any genus is straightforward.) Consider a charged chiral multiplet $\Phi_i$ of $R$-charge $r_i$ and gauge charges $\rho_i$ (the weights of the representation $\FR_i$), in a particular flux sector \eqref{flux sector}, together with its charge conjugate multiplet $\t\Phi_i$. Let us define:
\be
{\bf r}_{\rho_i} = r_i - \rho_i(k)~.
\ee
The scalar field component $\phi^{(\rho_i)}$ is a section of a line bundle $\CO(-{\bf r}_{\rho_i})$ over $\mathbb{P}^1$, with first Chern class $-{\bf r}_{\rho_i} $. Its zero-modes are holomorphic sections of $\CO(-{\bf r}_{\rho_i})$, which exist if and only if ${\bf r}_{\rho_i}\leq 0$. The analysis for the other chiral multiplet fields $\CC_{\b 1}, \t\phi$ and  $\t\CB$ is similar.
For each weight $\rho_i$ of the representation $\FR_i$, one has the following zero-modes:
\be\label{zero modes count chiral}
\Phi_{\rho_i} \rightarrow \begin{cases}
-{\bf r}_{\rho_i} +1 \quad\, {\rm zero\mbox{-}modes\; of}\; (\phi~, \t\phi~, \t\CB)^{(\rho_i)} &\; {\rm if}\;\;  {\bf r}_{\rho_i}\leq 0~,\cr
{\bf r}_{\rho_i} -1   \quad\quad {\rm zero\mbox{-}modes\; of} \; \CC_{\b 1}^{(\rho_i)}  &\; {\rm if}\;\;  {\bf r}_{\rho_i} \geq 1~.\end{cases}
\ee
Similarly, for a Fermi multiplet $\Lambda_I$ and its charge conjugate $\t\Lambda_I$, with $R$-charge $r_I$ and gauge representation $\FR_I$,  one finds:
\be\label{zero modes count Fermi}
\Lambda_{\rho_I} \rightarrow \begin{cases}
{\bf r}_{\rho_I} \quad\quad {\rm zero\mbox{-}modes\; of}\; \t\Lambda_I &\; {\rm if}\;\;  {\bf r}_{\rho_I} \geq 1~,\cr
-{\bf r}_{\rho_I} \quad \,{\rm zero\mbox{-}modes\; of}\; \Lambda_I &\; {\rm if}\;\;  {\bf r}_{\rho_I} \leq 0~,
\end{cases}
\ee
where we defined ${\bf r}_{\rho_I} = r_I - \rho_I(k)$. The zero-modes  \eqref{zero modes count chiral}-\eqref{zero modes count Fermi} are present if we turn off all interactions, while most of then are generally lifted by the gauge and $\CE_I$ couplings. In addition, we also have $\rk$ gaugino zero modes $\t\lambda_a$ ($a=1, \cdots, \rk$)  from the vector multiplet \eqref{vec mult02}.

\section{$A/2$-twisted  GLSM with an $\CN{=}(2,2)$ locus}\label{sec: loc in GLSM 1}

In this section, we consider an $\CN{=}(0,2)$ GLSM with an $\CN{=}(2,2)$ locus. In terms of $\CN{=}(0,2)$ multiplets, the theory contains  a $\Fg$-valued vector multiplet, a chiral multiplet $\Sigma$ in the adjoint representation of $\Fg$, and pairs of chiral and Fermi multiplets $(\Phi_i, \Lambda_I)$, with  $i=I$,  transforming in  representations $\FR_i$ of $\Fg$.  On the $\CN{=}(2,2)$ locus, the $\CE_I$ and $J_I$ potentials read
\be\label{E and J on 22}
\CE_I = \Sigma \Phi_i~,  \qquad \qquad J_I =  \d_{\Phi_i} W(\Phi)~,\qquad\qquad (I=i)~,
\ee
where $\Sigma$ acts on $\Phi_i$ in the representation $\FR_i$, and $W$ is the $\CN{=}(2,2)$ superpotential.  More generally, any properly gauge-covariant holomorphic functions $\CE_I$, $J_I$ are allowed as long as \eqref{JE constraint} is satisfied. (On the $\CN{=}(2,2)$ locus,   $\CE^I J_I=0$  follows from the gauge invariance of $W$.)

We choose to assign the following $R$-charges to the matter fields:~\footnote{ In the examples we will consider,  the $R$-charges  $r_i$ will all be either $0$ or $2$.}
\be
R[\Sigma]=0~,\qquad  R[\Phi_i]=r_i ~,\qquad R[\Lambda_i]=r_i-1~, \qquad r_i \in \Z~.
\ee
 This assignment automatically satisfies \eqref{R nonanomalous}. The corresponding curved-space theory realizes  the so-called $A/2$-twist,  generalizing the $A$-twist off the $\CN{=}(2,2)$ locus. 
The potential functions $\CE_I$ and $J_I$ must have $R$-charges $r_i$ and $2- r_i$, respectively. On the $\CN{=}(2,2)$ locus, there also exists an axial-like $R$-symmetry $U(1)_{\rm ax}$ at the classical level. 
In $\CN{=}(0,2)$ language, it corresponds to an alternative $R$-charge assignment 
\be
R_{\rm ax}[\Sigma]=2~, \qquad R_{\rm ax}[\Phi_i]=0~, \qquad R_{\rm ax}[\Lambda_i]=1~. 
\ee
We restrict ourselves to  theories that preserve that $R_{\rm ax}$ off the  $\CN{=}(2,2)$ locus as well. This means that $\CE_I$ remains linear in $\Sigma$ while $J_I$ cannot depend on $\Sigma$ at all. The $A/2$-twisted supercharge $\t\CQ_{(A/2)}$ has $R_{\rm ax}$-charge $1$.  Note that $R_{\rm ax}$ is generally anomalous at one-loop.

 We would like to compute the correlation functions
\be\label{CO gen sigma 0}
\left\langle \CO(\sigma) \right\rangle^{(A/2)}_{\mathbb{P}^1}
\ee
in the $A/2$-twisted theory on the sphere, where $\CO(\sigma)$ is any gauge-invariant polynomial in the scalar $\sigma$ from the chiral multiplet $\Sigma$. These are the simplest operators in the $A/2$-type pseudo-chiral ring. The presence of the $R_{\rm ax}$ symmetry leads to simple selections rules for \eqref{CO gen sigma 0}.  
The gauge anomaly of $R_{\rm ax}$ assigns the charge
\be\label{Rax gauge anom}
R_{\rm ax}[q_A]= 2 b_0^A~, 
\ee
to the Abelian gauge coupling \eqref{def qA}, where  $b_0^A$ is the FI parameter $\beta$-function coefficient \eqref{beta function xi}.
Moreover, $R_{\rm ax}$ suffers from a ``gravitational'' anomaly upon twisting. Due to the presence of zero-modes on the sphere, the path integral measure picks up a non-zero $R_{\rm ax}$-charge:
\be\label{Agrav Ahalf}
R_{\rm ax}[Z^{(A/2)}_{\mathbb{P}^1}] = - 2  d_{\rm grav} ~, \qquad  \qquad d_{\rm grav} = -{\dim}({\mathfrak g}) - \sum_i (r_i-1) {\dim}({\mathfrak R}_i)~.
\ee
Therefore, the standard `ghost number' selection rules of the $A$-model remain valid away from the $(2,2)$ locus. 

We would like to compute the $A/2$-twisted correlation functions \eqref{CO gen sigma 0} by supersymmetric localization. As we will show, the recent results of \cite{Benini:2015noa, Closset:2015rna} for $A$-twisted $\CN{=}(2,2)$ correlation functions can be extended to this case, provided some genericity condition is satisfied.

\subsection{The $\CN{=}(0,2)$ Coulomb branch}\label{subsec: Coulomb branch}
Consider the Coulomb branch consisting of diagonal VEVs for $\sigma$:
\be
\sigma ={\rm diag }(\sigma_a)~,\quad \qquad a=1, \cdots, \rk~,
\ee
and similarly for the charge-conjugate field $\t\sigma$. The Coulomb branch has the form $\fM \cong \mathfrak{h}_\bC /W$, with $\Fh$ the Cartan subalgebra of $\Fg$ and $W$ the Weyl group of $\GG$. Let us also denote by $\t\fM\cong \mathfrak{h}_\bC\cong  \C^\rk$ the covering space of $\fM$.
At a generic point on $\t\fM$, the gauge group is Higgsed to its Cartan subgroup $\GH$,
\be\label{higgsing GtoH}
\GG\rightarrow \GH = \prod_{a=1}^{{\rm rank}(G)} U(1)_a~,
\ee
with algebra $\Fh$ (up to the Weyl group).
Consider the holomorphic potentials $\CE_I =  \CE_I(\sigma, \phi)$, linear in $\sigma$, of $R$-charge $r_i$ (with $I=i$), which  transform in the same representations $\FR_I$ of $\Fg$ as $\Lambda_I$. 
Here and in the rest of this section, we identify the indices $i=I, j=J$, etc.
On the Coulomb branch, we have
\be\label{EI pot on CB}
\CE_I = \sigma_a \, E^a_I(\phi)~,
\ee
for some holomorphic functions $E^a_I(\phi)~$, and the matter multiplets $\Phi_I, \Lambda_I$ acquire masses
\be\label{mass matrix MIJ}
M_{IJ} = \d_J \CE_I  \big|_{\phi=0} =  \sigma_a \, \d_J E^a_I\big|_{\phi=0}~.
\ee
Note that \eqref{mass matrix MIJ} transforms in the representation $\FR_I \otimes \b\FR_J$ of $\Fg$.
 Gauge- and $U(1)_R$-invariance implies that the mass matrix \eqref{mass matrix MIJ} is  block-diagonal (up to a relabeling of the indices), with each block consisting of fields transforming in the same gauge representation and having the same $R$-charge. Let us denote by $\gamma= \{I_\gamma \} \subset \{ I\}$ the subset of indices corresponding to each of these blocks, so that we can partition the indices as $\{I\} = \cup_\gamma \{I_\gamma \} $, and let   $\FR_\gamma=\FR_{I_\gamma}$ be the corresponding gauge representations. We also denote by $r_\gamma$ the corresponding $R$-charge.  (That is, the chiral and Fermi multiplets $\Phi_{I_\gamma}$ and $\Lambda_{I_\gamma}$ have $R$-charges $r_\gamma$ and $r_\gamma-1$, respectively.) 
Each block is  diagonal in representation space, and we introduce the notation:
\be\label{def Mgamma}
 {M_{I_\gamma J_\gamma\, }}^{\rho_\gamma\rho'_\gamma} = \delta^{\rho_\gamma\rho'_\gamma} \, \left(M_{(\gamma, \, \rho_\gamma)}\right)_{I_\gamma J_\gamma}~, \ee
for each block. In  \eqref{def Mgamma}, $\rho_\gamma, \rho'_\gamma$ are indices running over the weights of the representation $\FR_\gamma$.  We also write
\be\label{def Mgamma ii}
\det M_{(\gamma, \,\rho_\gamma)} = \det_{I_\gamma J_\gamma} \left(M_{(\gamma, \, \rho_\gamma)}\right)_{I_\gamma J_\gamma}~.
\ee
In the following, we shall assume that
\be\label{gener cond on masses i}
\det M_{(\gamma, \,\rho_\gamma)}  \neq 0~, \qquad \forall (\gamma, \, \rho_\gamma)~,
\ee 
at any {\it generic} point on the Coulomb branch.
 This ensures that all the matter fields  are massive on $\t\fM$ except at special loci of positive codimension. In particular, the condition \eqref{gener cond on masses i} rules out theories with $\CE_I=0$.

At a generic point on the Coulomb branch, we  can therefore integrate out the matter fields to obtain an effective $J_a$-potential:
\be\label{Ja effective}
J_{\eff}^a = \tau^a-{1\ov 2\pi i} \sum_\gamma \sum_{\rho_\gamma\in \FR_\gamma} \rho_\gamma^a \log\left(\det M_{(\gamma, \,\rho_\gamma)}\right) -\half \sum_{\alpha>0}\alpha^a~,
\ee
where $\rho_\gamma$ are the weights of $\FR_\gamma$ and $\alpha$ are the positive simples roots of $\Fg$. The classical couplings $(\tau^a)\in \Fh_\C^\ast$ are the complexified parameters of the effective theory, which are obtained  from the  parameters $\tau^A$ by embedding  the central sub-algebra $\mathfrak{c}_\bC^* \subset \mathfrak{h}_\bC^* \subset \mathfrak{g}_\bC^*$ of the dual of  $\mathfrak{g}$ into $\mathfrak{h}_\bC^*$.
The second  contribution to \eqref{Ja effective}  arises from integrating out the chiral and Fermi multiplets \cite{McOrist:2007kp}, and the last term is the contribution from the  $W$-bosons multiplets.  From \eqref{Ja effective}, we read off the effective FI parameter on the Coulomb branch. In particular, we are interested in the effective FI parameter at infinity on  $\t \fM$. Denoting by $R$ the overall radius of $\t \fM\cong \C^r$, we define:
\be\label{def xiUV}
\xi_\eff^\text{UV} = \xi + {1\over 2 \pi} b_0 \, \lim_{R \ra \infty} \log R~,\quad\qquad \quad b_0= \sum_i \sum_{\rho_i\in \FR_i} \rho_i~,
\ee
where $b_0 \in i \Fh^\ast$ is equivalent to  $(b_0^A)\in i \mathfrak{c}^*$ defined in \eqref{beta function xi}.

\subsection{Quantum sheaf cohomology relations}\label{subsec: QSC from CB}
Consider a GLSM such that all the chiral multiplets have vanishing $R$-charge.  In that case, the pseudo-chiral ring relations---or QSC relations---can be analyzed on the Coulomb branch, similarly to the $\CN{=}(2,2)$ case  \cite{McOrist:2007kp}. On $\t\fM$, these relations are encoded in the equations:
\be
\exp{(J^a_{\rm eff})} = 1~,\qquad \forall a~, 
\ee
which read:
\be\label{QSC general CB}
\prod_\gamma \prod_{\rho_\gamma\in \FR_\gamma} \left(\det M_{(\gamma, \,\rho_\gamma)}\right)^{\rho_\gamma^a } = (-1)^{\sum_{\alpha>0}\alpha^a}\, q_a~, \qquad \forall a~, 
\ee
with $q_a= e^{2\pi i \tau^a}$.
Let $S_{QSC}$ be the set of isolated solutions  $(\sigma_a)$ to  \eqref{QSC general CB} satisfying the additional constraint that they correspond to  points on the Coulomb branch with maximal Higgsing \eqref{higgsing GtoH}. (For instance, for a $U(N)$ gauge group this gives the additional conditions that $\sigma_a\neq \sigma_b$ if  $a\neq b$.)
The QSC relations are the relations $f(\sigma_0)=0$ satisfied by any element $\sigma_0$ of $S_{QSC}$. 

In the abelian case, the $\sigma_a$'s correspond to gauge invariant operators and \eqref{QSC general CB} are the QSC relations themselves \cite{McOrist:2007kp}. For non-abelian theories, it requires some additional ingenuity to extract the explicit gauge-invariant relations from the Coulomb branch description. 
We briefly discuss an important $U(N)$ example in section \ref{subsec: Grassmannian}.

\subsection{$A/2$-twisted correlation functions}\label{subsec: main result}
The correlation functions  \eqref{CO gen sigma 0}  can be computed explicitly as a sum over flux sectors on the sphere, with each summand given by a generalized Jeffrey-Kirwan (JK) residue on $\t\fM \cong \C^\rk$. We find:
\be\label{JK res formula 22 i}
\left\langle \CO(\sigma) \right\rangle^{(A/2)}_{\mathbb{P}^1}={(-1)^{N_*}\ov |W|} \sum_{k \in \Gamma_{\mathbf{G}^\vee}} q^k\,  {\text{JKG-Res}}\! \left[\eta \right]  \cZ_k^\oneloop(\sigma)\, \CO(\sigma)\,   d\sigma_1\wedge \cdots \wedge d\sigma_\rk ~,
\ee
with
\be\label{JK res formula 22 ii}
 \cZ_k^\oneloop(\sigma) = 
  (-1)^{\sum_{\alpha>0}(\alpha(k) +1)} \prod_{\alpha>0} \alpha(\sigma)^2 \, \prod_\gamma \prod_{\rho_\gamma \in\FR_\gamma} \left( \det M_{(\gamma, \, \rho_\gamma)}\right)^{r_\gamma - 1 - \rho_\gamma(k)}~.
\ee
Here and in the next subsection, we explain the notation used in \eqref{JK res formula 22 i}-\eqref{JK res formula 22 ii}.  The derivation of the formula is discussed in  subsection \ref{subsec: der}.

The overall factor in  \eqref{JK res formula 22 i} is Weyl symmetry factor, with $|W|$ the order of the Weyl group of $\GG$. The sign factor $(-1)^{N_*}$ is a sign ambiguity. In the examples we shall consider with chiral multiplets of $R$-charges $0$ and $2$ only, we should take $N_*$ to be the number of chiral multiplets of $R$-charge $2$ \cite{Morrison:1994fr, Closset:2015rna}.

The sum in \eqref{JK res formula 22 i} is over the GNO-quantized \cite{Goddard:1976qe} magnetic fluxes
$k \in \Gamma_{\mathbf{G}^\vee} \subset i\mathfrak{h}$.
The integral lattice $\Gamma_{\mathbf{G}^\vee} \cong \bZ^\rk$  can be obtained from $\Gamma_\mathbf{G}$, the weight lattice of electric charges of $\mathbf{G}$ within the vector space $i\mathfrak{h}^*$, by \cite{Englert:1976ng,Kapustin:2005py}
\be\label{flux lattice}
\Gamma_{\mathbf{G}^\vee} = \setcond{k}{\rho(k) \in \bZ ~~ \forall \rho \in \Gamma_\mathbf{G}}~,
\ee
where $\rho(k)= \sum_a\rho^a k_a$ is given by the canonical pairing of the dual vector spaces.  Let us also introduce the notation $\vec k \in \bZ^n$ to denote the fluxes in the free part \eqref{U1A factors}  of the center of  $\mathbf{G}$. We define
\be
q^k \equiv \exp(2 \pi i \sum_{A=1}^n (\vec \tau)_A (\vec k)_A) =  \exp(2 \pi i \tau(k))~.
\label{vqvK0}
\ee
Here $\vec \tau \in \bC^n$ denotes the complexified FI parameter, while $\tau$ is the same FI parameter viewed as an element of $\mathfrak{h}_\bC^*$.

Each summand in \eqref{JK res formula 22 i} is given by a  (conjectured) generalization of the JK residue \cite{1993alg.geom..7001J, 1999math......3178B, szenes-vergne}, called the JKG residue, upon which we elaborate shortly. That residue depends on the argument $\eta \in \mathfrak{h}_\bC^*$  in \eqref{JK res formula 22 i}. In this work, we take 
\be\label{eta equal xi}
\eta = \xi_\eff^\text{UV}~, 
\ee
 the effective FI parameter in the UV defined in \eqref{def xiUV}. With this choice, the JKG residue is a local operation at the origin of the Coulomb branch (or at finite distance on the Coulomb branch) because all boundary terms---the potential contributions from infinity on $\t \fM$---vanish \cite{Jia:2014ffa, Closset:2015rna}. (A different choice of $\eta$ would require a careful treatment of these boundary terms, but one can always choose \eqref{eta equal xi}---$\eta$ is an auxiliary parameter in the derivation, which cannot affect the physical result. The only restriction in the use of \eqref{eta equal xi} is that $\eta$ should not lie on a chamber wall in FI parameter space.)

The integrand in \eqref{JK res formula 22 i} is a meromorphic $\rk$-form on $\t\fM \cong \C^\rk$. The expression \eqref{JK res formula 22 ii} is the contribution from the massive fields on the Coulomb branch. The first product in \eqref{JK res formula 22 ii} runs over all the positive simple roots $\alpha>0$ of $\Fg$ and corresponds to the $W$-bosons. The second product in \eqref{JK res formula 22 ii} is the contribution from the matter multiplets $\Phi_I, \Lambda_I$, with the partition of indices $\{I\} = \cup_\gamma \{I_\gamma \} $ as explained above \eqref{def Mgamma}, and another product over all the weights $\rho_\gamma$ of the  representation $\FR_\gamma$ of $\Fg$, for each $\gamma$. The polynomials   $\det M_{(\gamma,\, \rho_\gamma)}$ were defined in \eqref{def Mgamma}-\eqref{def Mgamma ii}.

\subsection{The Jeffrey-Kirwan-Grothendieck residue}\label{subsec: JKG def}
Let us introduce the collective label $\cI_\gamma = (\gamma, \rho_\gamma)$ for the field components in each block $\gamma$.
In any given flux sector, the integrand in \eqref{JK res formula 22 i} is a meromorphic $(r,0)$-form  on%
~\footnote{Here and in the rest of this section, we often write $r=\rk$ to avoid clutter.} 
 $\t \fM\cong \Fh_\C  \cong \C^r$ with potential singularities at:
\be\label{potential singularities}
\cup_\gamma \CH_{\cI_\gamma}\subset \C^r~,  \qquad\qquad \CH_{\cI_\gamma} \cong \{\sigma \in \C^r\,  | \, \det M_{\cI_\gamma} =0 \}~.
\ee
Each $\CH_{\cI_\gamma}$ is a divisor (codimension-one subvariety~\footnote{We use the terms `divisor' and `codimension-one variety' interchangeably. That is, all our divisors are effective.}) of $\C^r$ and all  these divisors intersect at  $\sigma=0$. Let us denote by
\be\label{def Pgamma}
P_{\cI_\gamma}(\sigma)= \det M_{\cI_\gamma}(\sigma) \in \C[\sigma_1, \cdots, \sigma_r]
\ee
the homogeneous polynomials of degree $d_\gamma$ associated to \eqref{potential singularities}. (For each $\gamma$, every $P_{\cI_\gamma}$ has the same degree.)   To each $P_{\cI_\gamma}$, we associate the charge vector $Q_{\cI_\gamma}\in i \Fh^\ast$, which is the $U(1)^r$ gauge charge of the field component $\cI_\gamma$ under the Cartan subalgebra $\Fh$---that is:
\be
Q_{\cI_\gamma}^a = \rho^a_\gamma~, 
\ee
if $\cI_\gamma = (\gamma, \rho_\gamma)$. In any flux sector with flux $k$, the actual singularities consist of the subset  of the potentials singularities \eqref{potential singularities} at $P_{\cI_\gamma}$ such that
\be\label{cond to have sing}
 \rho_\gamma(k)-r_\gamma\geq 0~.
\ee
We shall {\it assume} that, in any given flux sector, the set of  charge vectors  ${\bf Q}\subset\{ Q_{\cI_\gamma}\}$ associated to the actual singularities is {\it projective}---that is, the vectors ${\bf Q}$ are contained within a half-space of $i\Fh^\ast$.  Note that a non-projective ${\bf Q}$  signals the presence of dangerous gauge invariant operators which may take an arbitrarily large VEV \cite{Closset:2015rna}.  One can sometimes render a non-projective singularity projective by  turning on some twisted masses of the type considered in section \ref{subsec: mass def} below, effectively splitting the singularity.

We would like to define the  ``Jeffrey-Kirwan-Grothendieck'' (JKG) residue as a simple generalization of the Jeffrey-Kirwan residue. Let us first recall the definition of the Grothendieck residue \cite{GriffithsHarris}  specialized to our case.  Given $r$ homogeneous polynomials $P_b$, $b=1, \cdots, r$,   in $\C[\sigma_1, \cdots, \sigma_r]$, of degrees $d_b$, such that $P_{1}=\cdots=P_{r} =0$ if and only if $\sigma_1= \cdots = \sigma_r=0$, let us define a $(r,0)$-form on $\C^r$:
\be\label{def omegaP}
\omega^{(P)} = {d\sigma_1\wedge\cdots \wedge d\sigma_r\ov  P_{1}(\sigma)\cdots  P_{r}(\sigma)}~.
\ee
Let $D_b$ be the divisor in $\C^r$ corresponding to $P_b=0$, and let $D_P= \cup_b D_b$. The form \eqref{def omegaP} is holomorphic on $\C^r\backslash D_P$.
The Grothendieck residue of $f \, \omega^{(P)}$ at $\sigma=0$, with $f=f(\sigma)$ any holomorphic function, is given by:
\be\label{def resG}
{\rm Res}_{(0)} \, f \, \omega^{(P)} ={1\over (2\pi i)^r} \oint_{\Gamma_{\epsilon}}   f \, \omega^{(P)}~,
\ee
with a real $r$-dimensional contour:
\be
\Gamma_{\epsilon} =\left\{\sigma \in \C^r\,  \big| \, |P_{l_b}| =\epsilon_b~, \; b=1, \cdots, r \right\}~,
\ee
oriented by $d(\arg(P_{l_1}))\wedge \cdots\wedge d(\arg(P_{l_r}))\geq 0$, with $\epsilon_b >0$, $\forall b$. The  residue \eqref{def resG}   only depends on the homology class of $\Gamma_{\epsilon}$ in $H_n(\C^r \backslash  D_{P})$. Note that, if $f$ is an homogenous polynomial of degree $d_0$, the residue \eqref{def resG}  vanishes unless $d_0=\sum_{b=1}^r (d_b-1)$.
Useful properties of  the Grothendieck residue are reviewed in appendix \ref{app: residues}.

Consider  an arrangement of $s\geq r$ distinct irreducible divisors $\CH_{\cI_\gamma} \cong \{\sigma \, |\,P_{\cI_\gamma}=0\}$ of $\Fh_\C \cong \C^r$,  intersecting at $\sigma=0$, and  denote by $D_{\bf P}$ their union. To each $\cI_\gamma$ is associated the charge $Q_{\cI_\gamma} \in i \Fh^\ast$. We denote this data by:
\be
{\bf P}= \{ P_{\cI_\gamma} \}~,\qquad\qquad {\bf Q}= \{ Q_{\cI_\gamma }\}~,
\ee
were ${\bf Q}$ is assumed projective. 
Let $R_{\bf P}$ be the space of rational holomorphic $(r,0)$-forms with poles on $D_{\bf P}$, and let $S_{\bf P}\subset R_{\bf P}$ be the linear span of
\be\label{def omegasSP}
\omega_{S, P_0} = d\sigma_1\wedge\cdots \wedge d\sigma_r \prod_{P_b \in P_S} {P_0 \over P_b}~, 
\ee
where $P_S= \{ P_1, \cdots, P_r\}\subset {\bf P}$ denotes any subset of $r$ distinct polynomials in ${\bf P}$ associated to $r$ distinct charges $Q_S= \{ Q_1, \cdots, Q_r\} \subset {\bf Q}$, while $P_0$ is any homogeneous polynomial of degree $d_0= \sum_{b=1}^r (d_b-1)$, with $d_b$ the degree of $P_b$. 
The JKG-residue on $S_{\bf P}$ is defined by 
\be\label{def JKG 1}
  {\text{JKG-Res}}\! \left[\eta \right] \,  \omega_S=
   \left\{ 
\begin{array}{ll}  \sign\left(\det(Q_S)\right) \,  {\rm Res}_{(0)} \,  \omega_S   &\qquad {\rm if}\quad \eta\in \text{Cone}(Q_S)~, \\
 0 &\qquad {\rm if}\quad \eta\notin \text{Cone}(Q_S)~,
\end{array} \right.
\ee
in terms of a vector $\eta\in \Fh^\ast$. Here,  $ \text{Cone}(Q_S)$ denotes the positive span of the $r$ linearly-independent vectors $Q_S$ in $\Fh^\ast$. We further {\it conjecture} that there exists a canonical projection $R_{\bf P}\rightarrow S_{\bf P}$, so that the JKG residue is defined on $R_{\bf P}$ through \eqref{def JKG 1} by composition, similarly to the JK residue defined in \cite{1999math......3178B}.

The contour integral in  \eqref{JK res formula 22 i} is a  JKG-residue at the origin, with the vector $\eta$ given by \eqref{eta equal xi}. Oftentimes,  one can find the correct JKG contour by considering small deformations off the $\CN{=}(2,2)$ locus.
On the $\CN{=}(2,2)$ locus,  the divisors  $\CH_{\cI_\gamma}$ are hyperplanes  orthogonal to $Q_{\cI_\gamma}$, with
\be
P_{\cI_\gamma}=\left(Q_{\cI_\gamma}(\sigma)\right)^{d_{\cI_\gamma}}~,
\ee
and  the JKG-residue reduces to an ordinary Jeffrey-Kirwan residue, reproducing previous results for the $A$-twisted GLSM \cite{Benini:2015noa,Closset:2015rna}.

\subsection{Derivation of the JKG residue formula}\label{subsec: der}
In this subsection, we sketch a derivation of the residue formula \eqref{JK res formula 22 i}, closely following previous works \cite{Benini:2013xpa, Jia:2014ffa, Closset:2015rna}, to which we refer for more details. We shall leave one important technical step---the proper cell decomposition of the Coulomb branch---as a conjecture. More generally, we would like to stress that the JKG residue has not yet been defined satisfactorily at the mathematical level.  We hope that the present work will motivate further investigation of the subject.

\subsubsection{Generalities}
We use the kinetic terms of section \ref{subsec: kin} in the localizing action:
\be
\SL_{\rm loc} = {1\ov e^2}\left( \SL_{YM}+ \SL_{\t\Sigma\Sigma}\right) + {1\ov g^2}\left( \SL_{\t\Phi\Phi} + \SL_{\t\Lambda\Lambda}\right)~,
\ee
with $e$ and $g$ some dimensionless parameters that we can take arbitrarily small. With the standard reality condition $\t\sigma = \b \sigma$, the kinetic term for the chiral multiplet $\Sigma$ localizes to~\footnote{More precisely, we performed a field redefinition of the auxiliary field $D$ so that the Lagrangian $\SL_{YM}+ \SL_{\t\Sigma\Sigma}$ match the  $\CN{=}(2,2)$ SYM Lagrangian. That introduces a term $[\sigma, \t\sigma]^2$ in the action. See \cite{Benini:2013xpa} for a similar discussion.}
\be
\d_\mu \sigma = 0~, \qquad \qquad [\sigma, \t\sigma]= 0~.
\ee
We therefore localize onto the Coulomb branch discussed in subsection \ref{subsec: Coulomb branch}. We also have a sum over gauge fluxes,
\be
k ={1\ov 2 \pi} \int_{\mathbb{P}^1} d a~,
\ee
with $k$ in the flux lattice \eqref{flux lattice}. Let us define 
\be
\h D = - i \left(D- 2 i f_{1\b1}\right)~,
\ee
with $\h D$ a real field corresponding to fluctuations around the supersymmetric value $\h D=0$, in any topological sector.
At a generic points on the Coulomb branch, all the other matter field are massive, while for special values of $\sigma$ corresponding to 
\be
P_{\cI_\gamma}(\sigma)=0~,
\ee
with $P_{\cI_\gamma}$ defined in \eqref{def Pgamma}, we have additional bosonic zero modes and the localized path integral would be singular. To regularize these singularities, it is useful to keep the constant mode of $\h D$  in intermediate computations \cite{Benini:2013xpa}. 

We also have the fermionic zero modes $\t\lambda$ from the Coulomb branch vector multiplets, and the fermionic zero modes $\t\CB^\Sigma$ from $\Sigma$---corresponding to \eqref{zero modes count chiral} with ${\bf r}=0$. The path integral localizes to:
\be\label{Zglsm intermed}
Z_{\rm GLSM}= {1\ov |W|}\sum_k q^k \int \prod_{a}^\rk \left[d^2\sigma_a\,  d\h D_a \, d \t\lambda_a \, d\t\CB^\Sigma_a  \right] \, \CZ_k(\sigma, \t\sigma, \t\lambda, \t\CB^\Sigma, \h D)~,
\ee
where $\CZ_k(\sigma, \t\sigma, \t\lambda, \t\CB^\Sigma, \h D)$ is the result of integrating out the matter fields and W-bosons in the supersymmetric background: 
\be
\CV_0=(\t \lambda_a~, \h D_a)~, \qquad \Sigma_0 =(\sigma_a~, \t\sigma_a~, \t\CB^\Sigma_a)~.
\ee
Supersymmetry implies the relation:
\be\label{susy rel Zk}
\delta \CZ_k = \left(\h D_a {\d\phantom{n}\ov \d \t\lambda_a} + \t\CB^\Sigma_a{\d\phantom{n}\ov \d \t\sigma_a}\right)\CZ_k=0~.
\ee
In the limit $e, g\rightarrow 0$, we have
\be\label{CZk def}
\CZ_k(\sigma, \t\sigma, \h D) \equiv \CZ_k(\sigma, \t\sigma, 0, 0, \h D) =\lim_{e\rightarrow 0}  e^{-S_0} \, \CZ_{k}^{\rm massive}(\sigma, \t\sigma, \h D) \,   \cZ_k^\oneloop(\sigma, \t\sigma, \h D)~.
\ee
Here, $e^{-S_0}$ is the classical contribution, with
\be
S_0 = {\rm vol}(S^2) \left({1\ov 2e^2} \h D^2 -\half \t\tau(\h D) \right)~,
\ee
(setting $e_0=1$ in \eqref{SYM lag}), while  $\CZ_{k}^{\rm massive}$ is the contribution from non-zero modes, which is trivial when $\h D=0$, and $\cZ_k^\oneloop$ is the zero-mode contribution, which reduces to \eqref{JK res formula 22 ii} when $\h D=0$. These one-loop contributions are derived and discussed in  appendix \ref{sec: oneloop det}.

The insertion of any pseudo-chiral operator $\CO(\sigma)$ does not modify the derivation. It simply corresponds to inserting the same factor $\CO(\sigma)$ with constant $\sigma$ in the integrand \eqref{Zglsm intermed}.

\subsubsection{The rank-one case}
Consider first the case of a rank-one gauge group. We  choose $\GG= U(1)$ for simplicity, but the generalization is straightforward. We have matter fields $\Phi_i, \Lambda_i$ with gauge charges $Q_i$ and $R$-charges $r_i$ and $r_i-1$, organized in blocks $\Phi_\gamma$. We have the one-loop contributions 
\be
 \CZ_{k}^{\rm massive}(\sigma, \t\sigma, \h D)= \prod_\gamma \prod_{\lambda_{(\gamma, k)}} {\det(\lambda_{(\gamma, k)}+ |M_\gamma|^2)\ov \det(\lambda_{(\gamma, k)}+ |M_\gamma|^2 + i Q_\gamma \h D)}
\ee 
with $\lambda_{(\gamma, k)}>0$,  and
\be
\cZ_k^\oneloop(\sigma, \t\sigma, \h D) = \prod_\gamma \cZ_{k, \gamma}^\oneloop
\ee
with
\be
\cZ_{k, \gamma}^\oneloop=\begin{cases}
(\det M_\gamma)^{r_\gamma -1 - Q_\gamma k} &\; {\rm if}\;\;  r_\gamma - Q_\gamma k \geq 1~,\cr
\left(\det \b M_\gamma \ov   \det\left( |M_\gamma|^2 + i Q_\gamma \h D\right) \right)^{1-r_\gamma + Q_\gamma k} &\; {\rm if}\;\; r_\gamma - Q_\gamma k < 1~,\end{cases}
\ee
from the zero modes. The singular  locus on the Coulomb branch corresponds to $\det M_\gamma=0$, for each $\gamma$. This is simply $\sigma=0$ in the present case, but it is useful to suppose that $\det M_\gamma$ has more general roots. (That can be achieved with twisted masses, as in section \ref{subsec: mass def} below.) 
In each flux sector, we remove a small neighborhood $\Delta_{\epsilon, k}$ of the singular locus, of size $\epsilon>0$, and we decompose this neighborhood as
\be
\Delta_{\epsilon, k}= \Delta_{\epsilon, k}^{(+)} \cup \Delta_{\epsilon, k}^{(-)}\cup \Delta_{\epsilon, k}^{(\infty)}~,
\ee
where $\Delta_{\epsilon, k}^{(\pm)} $ corresponds to the neighborhood of the singularities from the positively and negatively charged matter fields ($Q_\gamma >0$ and $Q_\gamma<0$, respectively), as well as the neighborhood of $\sigma=\infty$. We assume that our theory is such that we can always separate the singularities from positively and negatively charged fields, for any given $k$. (Such singularities are `projective singularities' in the sense defined below \eqref{cond to have sing}.)

Using \eqref{susy rel Zk}, one can perform the integration over the fermionic zero modes in \eqref{Zglsm intermed}, to obtain:
\be
Z_{\rm GLSM}= \sum_k q^k   \int_\Gamma  {d\h D\ov \h D}  \oint_{\d\Delta_{\epsilon, k}} d\sigma\, \CZ_k(\sigma, \t\sigma,  \h D)~,
\ee
For each $\gamma$ block, the Hermitian matrix  $|M_\gamma|^2$ can be diagonalized with eigenvalues $m_\gamma^2 >0$. The absence of chiral multiplet tachyonic modes requires that
\be
{\rm Im}(Q_\gamma \h D) < m_\gamma^2~, \qquad \forall \gamma, \forall m_\gamma^2~.
\ee 
This determines the $\h D$ contour of integration $\Gamma$ exactly like in \cite{Closset:2015rna}. 
There is an important contribution from infinity, which is controlled by the effective FI parameter \eqref{def xiUV}. We have a twofold freedom in choosing $\Gamma$ (corresponding to the sign of $\eta$ in \eqref{def JKG 1}) and we can choose $\eta=\xi_\eff^\text{UV}$ so that  the contribution from $\d \Delta_{\epsilon, k}^{(\infty)}$ vanishes \cite{Jia:2014ffa, Closset:2015rna}.
In that case, performing the $\h D$ integral picks the contributions from $\d\Delta_{\epsilon, k}^{(+)}$ or $\d\Delta_{\epsilon, k}^{(-)}$ according to the sign of $\xi_\eff^\text{UV}$:
\be\label{Zglsm res rk1}
Z_{\rm GLSM}^{(+)}=  \sum_k q^k  \oint_{\d\Delta_{\epsilon, k}^{(+)}} d\sigma\, \CZ_k^\oneloop(\sigma)~, \quad\;
Z_{\rm GLSM}^{(-)}=-\sum_k q^k  \oint_{\d\Delta_{\epsilon, k}^{(-)}} d\sigma\, \CZ_k^\oneloop(\sigma)~.
\ee
The first equality corresponds to $\eta= \xi_\eff^\text{UV} >0$ and the second equality corresponds to $\eta=\xi_\eff^\text{UV}<0$. When $b_0=0$,   $\xi_\eff^\text{UV} $ can be tuned to be of either sign and the two formulas \eqref{Zglsm res rk1} are equal as formal series \cite{Closset:2015rna}. The result \eqref{Zglsm res rk1} can be written as the JKG residue  \eqref{def JKG 1}.

\subsubsection{The general case}
In the general case, one can perform the fermionic integral in \eqref{Zglsm intermed} explicitly to obtain:
\be\label{Zglsm intermed ii}
Z_{\rm GLSM}= {1\ov |W|}\sum_k q^k \int \prod_{a}^\rk \left[d\sigma_a\,d\t\sigma_a \, d\h D_a \right] \,  \det_{ab}(h_{ab})\, \CZ_k(\sigma, \t\sigma, \h D)~,
\ee
with $h_{ab}$ a two-tensor on $\t\fM$ that satisfies
\be\label{prop of h}
\d_{\t \sigma_a} h_{bc}- \d_{\t \sigma_c} h_{ba}=0~, \qquad \quad
\d_{\t \sigma_a}  \CZ_k(\sigma, \t\sigma, \h D) = \h D^b h_{b a} \, \CZ_k(\sigma, \t\sigma, \h D)~,
\ee
with $\CZ_k(\sigma, \t\sigma, \h D)$ given in \eqref{CZk def}. The only difference with the discussion in \cite{Closset:2015rna} is that $h_{ab}$ need not be symmetric. One way to motivate this result is to note that the low-energy effective action on the Coulomb branch should take the form
\be
S_{\rm eff} \propto - \h D^a \t J_a^{\rm eff} + \t\lambda^a {\d \t J_a^{\rm eff} \ov \d \sigma_b} \t\CB^\Sigma_b~,
\ee
with $\t J_a^{\rm eff}$ the anti-holomorphic effective superpotential. Therefore, we have $h_{ab} ={\d \t J_a \ov \d \sigma_b} $ and the properties \eqref{prop of h} follow. More generally, the $h_{ab}$ in \eqref{Zglsm intermed ii} may depend on $\h D_a$ but the above properties are preserved and follow from supersymmetry. We may define a form
\be
\nu(V) = V^a h_{ab} d\t\sigma^b
\ee
for any $V$ valued in $\Fh_\C$, in terms of which \eqref{prop of h} reads
\be
\b\d \nu=0~, \qquad \quad \b\d \CZ_k = \nu(D) \CZ_k~,
\ee
with $\b\d$ the Dolbeault operator on $\t\fM$.   In any flux sector, we define $\Delta_{\epsilon, k}$ to be the union of the small neighborhoods of size $\epsilon$ around the divisors $\CH_{\cI_\gamma}$ in  \eqref{potential singularities} such that \eqref{cond to have sing} holds, and of the neighborhood of $\sigma=\infty$. We have
\be
Z_{\rm GLSM}={1\ov |W|}  \lim_{e, \epsilon\rightarrow 0} \sum_k  q^k \int_{\Gamma \ltimes \tfM \setminus \Delta_{\epsilon, k}} \mu_{(k)}~,
\ee
where $r=\rk$ and $\mu_{(k)} $ is a top-form:
\be
\mu_{(k)} = {1\ov r!} \CZ_k(\sigma, \t\sigma, \h D) \, d^r\sigma \wedge \nu(d\h D)^{\wedge r}~.
\ee
From here onward, one may follow \cite{Closset:2015rna} almost verbatim.
 The main difficulty lies in dealing with the boundaries of $\Delta_{\epsilon, k}$, the tubular neighborhood of the singular locus that should be excised from $\t\fM$.   We conjecture that a sufficiently good cell decomposition exists, such that the manipulations of \cite{Benini:2013xpa, Jia:2014ffa, Closset:2015rna} can be repeated while replacing the singular hyperplanes by singular divisors. This would establish the JKG residue prescription in the regular case, that is when the number $s$ of singular divisors equals $r$. (The prescription for the non-regular case, $s>r$, is a further conjecture, motivated by examples.)

\subsection{Comparison to previous results}
It is convenient to rewrite \eqref{JK res formula 22 i} as:
\be\label{JK res formula 22 i BIS}
\left\langle \CO(\sigma) \right\rangle^{(A/2)}_{\mathbb{P}^1}={(-1)^{N_*}\ov |W|} \sum_{k \in \Gamma_{\mathbf{G}^\vee}}   {\text{JKG-Res}}\! \left[\eta \right]  e^{2\pi i J_{\rm eff}(k) }  \cZ_k^\oneloop(\sigma)\, \CO(\sigma)\,   \,d^{\rk}\sigma~,
\ee
where $J_{\rm eff}(k)= J^a_{\rm eff} k_a$, with $J^a_{\eff}$ the effective $J_a$-potential defined in \eqref{Ja effective}, and
\be\label{JK res formula 22 ii BIS}
 \cZ_0^\oneloop(\sigma) = 
  (-1)^{\half {\rm dim}(\Fg/\Fh)} \prod_{\alpha>0} \alpha(\sigma)^2 \, \prod_\gamma \prod_{\rho_\gamma \in\FR_\gamma} \left( \det M_{(\gamma, \, \rho_\gamma)}\right)^{r_\gamma - 1}~.
\ee
Following \cite{Closset:2015rna}, let us assume that the integrand of  \eqref{JK res formula 22 i BIS} is such that the contributing fluxes  all lie within a discrete cone $\Lambda \subset \Gamma_{\mathbf{G}^\vee} $, defined by 
\be
\Lambda = \setcond{k}{k=\sum_A {n_A \kappa^A } + r^{(0)},   \quad
n_A \in \bZ_{\geq 0}}
\ee
for some $r^{(0)} \in \Gamma_{\mathbf{G}^\vee}$, with $\kappa^A$ ($A=1, \cdots, \rk$) a basis of $\Gamma_{\mathbf{G}^\vee}$, and such that, for every contributing flux, the JKG residue includes all the poles.  In such a situation, one can perform the sum over fluxes to obtain 
\be\label{JK res formula 22 i TER}
\left\langle \CO(\sigma) \right\rangle^{(A/2)}_{\mathbb{P}^1}=
{(-1)^{N_*} \over |W|}\oint_{\p \tfM}
\left( \prod_{a=1}^\rk {d\sigma_a \over 2\pi i} \right)
{e^{2 \pi i r^{(0)}_a J^a_{\rm eff}}
\over \prod_{A=1}^\rk ( 1 - e^{2\pi i \kappa^A_a J^a_{\rm eff}})}
\cZ_0^{\oneloop}(\sigma)\,  \CO\left(\sigma\right)~.
\ee
 Here the contour is the $\rk$-torus at infinity.  
Furthermore, if the theory has isolated massive Coulomb vacua, we can perform the integral  explicitly. Let us denote:
\be
P= \left \{ \sigma_P \,\big|\,  e^{2\pi i \kappa^A_a J^a_{\rm eff}(\sigma_P)}=1
~~\text{for all}~A=1,\cdots,\rk~\right\}~.
\ee
From \eqref{JK res formula 22 i TER}, we obtain
\be\label{corr sum over Coulomb vac}
\left\langle \CO(\sigma) \right\rangle^{(A/2)}_{\mathbb{P}^1}= 
{(-1)^{N_*}\over |W|} {1\over (-2\pi i)^\rk} \sum_{\sigma_P \in P}  { \cZ_0^\oneloop (\sigma_P)\,  \CO\left(\sigma_P\right)\over  \det_{AB}\left(\kappa^A_a \kappa^B_b (\d_{\sigma_b} J_{\rm eff}^a(\sigma_P))\right)}~,
\ee
This result was first obtained in \cite{McOrist:2007kp} for $\GG$ abelian. 
Another expression for the correlation functions was described in \cite{Lu:thesis, Lu:2015a}, were it was shown to be equivalent to the result of \cite{McOrist:2007kp}. Note that \eqref{corr sum over Coulomb vac} generalizes \cite{McOrist:2007kp} to the non-abelian case as well.

\subsection{Some properties of the  correlation functions}
\label{subsec: properties A2 corr}

The localization result \eqref{JK res formula 22 i}  renders  some interesting properties of the $A/2$-twisted correlation functions manifest, in the case of the GLSM with an $\CN{=}(2,2)$ locus that we are considering here. Specifically, we see that the correlations functions are independent of non-linear deformations of the $\CE_I$-potential, and that they are also completely independent of the superpotential $J_I$ except for the implied constraints on the $R$-charges of the chiral and Fermi multiplets. Note that these properties are not a direct consequence of $\CN{=}(0,2)$ supersymmetry---in particular, the corresponding couplings are not $\delta$-exact.


The $\CE_I$-potentials are linear in the field $\sigma$ but they can be of higher order in the other chiral multiplet scalars $\phi_I$,  as in \eqref{EI pot on CB}, if allowed by gauge invariance. The localization computation, however, only depends on the first-order terms in $\phi_I$ through the effective masses \eqref{mass matrix MIJ}, because the localization locus is simply $\phi_I=0$. Therefore, the $A/2$-twisted correlation functions \eqref{JK res formula 22 i} are independent of the non-linear terms in the $\CE_I$ potentials. 
This result was conjectured in \cite{McOrist:2008ji} for both quantum sheaf 
cohomology ring relations and correlation functions, proven rigorously for the
quantum sheaf cohomology of abelian models in 
\cite{Donagi:2011uz,Donagi:2011va} and argued
in \cite{melpriv} for correlation functions.  Here we derived the same
results  in greater generality.

It was also argued in \cite{McOrist:2008ji} that $A/2$-twisted GLSMs with a $(2,2)$ locus
 should be independent of the $J_I$ superpotential deformations. (The issue was later addressed in \cite{Garavuso:2013zoa}.)
This claim is rather striking from the point of view of the infrared NLSM, since it implies an analogue of the distinction between complex and K\"ahler structure moduli that exists for $\CN{=}(2,2)$ superconformal models. Our result proves this conjecture by explicit computation. We simply see that $J_I=0$ on the localization locus $\phi_I=0$, therefore the result is completely independent of the corresponding coupling constants. The only dependence on the $J_I$-superpotential is through the constraints that the presence of such terms impose on the allowed $R$-charges.

\section{Abelian examples}\label{sec: Abelian expl}

$\CN{=}(0,2)$ deformations of $\CN{=}(2,2)$ abelian GLSMs have been studied extensively in the literature. In particular, explicit results are known for the  correlation functions and for the quantum sheaf cohomology ring of models describing toric varieties with a deformed tangent bundle---see {\it e.g.}  \cite{McOrist:2008ji,Donagi:2011uz,Donagi:2011va}. In this section, we rederive some of those results using our localization formula, which simplify the computations considerably.

\subsection{The ${\mathbb P}^{N_f-1}$ model}
The tangent bundle of ${\mathbb P}^{N_f-1}$ can be defined by a short exact sequence:
\be
0 \: \longrightarrow \: {\cal O} \: \stackrel{*}{\longrightarrow} \:
{\cal O}(1)^{N_f} \: \longrightarrow \: T {\mathbb P}^{N_f-1} \: \longrightarrow
\: 0~,
\ee
where $*$ is given by multiplication by homogeneous coordinates. $T {\mathbb P}^{N_f-1}$ admits no holomorphic deformations. The corresponding GLSM consists of a $U(1)$ vector multiplet, one neutral chiral multiplet $\Sigma$, and $N_f$ chiral and Fermi multiplets $\Phi_I, \Lambda_I$ with gauge charge $Q=1$ and vanishing $R$-charge. The most general $\CE_I$ potential allowed is
\be
\CE_I = \sigma\, {A_I}^J \phi_I~,
\ee
with $A$ a constant $N_f\times N_f$ matrix. We take $A$ to be invertible so that the Coulomb branch of section \ref{sec: loc in GLSM 1} exists, which implies that $A$ can be set to unity by a field redefinition. In that case the model actually possesses $\CN{=}(2,2)$ supersymmetry. In is instructive, however, to consider an arbitrary invertible $A$ as a formal deformation. 

In this simple case, we have a single $\gamma$-block and the Coulomb branch mass matrix:
\be
{M_I}^J = \sigma \, {A_I}^J~.
\ee
The formula \eqref{JK res formula 22 i} gives:
\be  \label{eq:pN-corrfn}
\langle \sigma^n  \rangle  =
\sum_{k =0}^\infty  q^k \oint\frac{d \sigma}{2\pi i}
\frac{\sigma^{n}}{(\det M)^{k+1}}
=
\begin{cases} (\det A)^{-k-1} \, q^k  &\text{if } n = N(k+1)- 1~,\cr 
0 &\text{otherwise}. \end{cases}
\ee
In the first line, we used the fact that $\xi_\eff^\text{UV} \rightarrow  +\infty$ in this model, from which it follows that only the fluxes $k\geq 0$ contribute to the JKG residue. 
The result  \eqref{eq:pN-corrfn}  differs from the $\CN{=}(2,2)$ result by a rescaling of $q$ to $(\det A)^{-1} q$, and by an overall factor of  $(\det A)^{-1}$ which could be reabsorbed in a local couterterm \eqref{local ctterm}. Note that the correlations functions are singular at $\det{A}\rightarrow 0$, which corresponds to the appearance of additional massless modes on the Coulomb branch. (For $A=0$, $\Sigma$ itself becomes free.)

\subsection{The ${\mathbb P}^1 \times {\mathbb P}^1$ model}
\label{sect:ex:p1xp1}

This is one of the simplest
examples of a toric variety with nontrivial tangent bundle deformations.
Consider the holomorphic bundle ${\bf E}$ over
${\mathbb P}^1 \times {\mathbb P}^1$  realized as a cokernel by the short exact sequence:
\be
0 \: \longrightarrow \: {\cal O}^2 \: \stackrel{*}{\longrightarrow} \:
{\cal O}(1,0)^2 \oplus {\cal O}(0,1)^2 \: \longrightarrow \: {\bf E}
\: \longrightarrow \: 0~, 
\ee
where
\be\label{map def E P1P1}
* \: = \: \left[ \begin{array}{cc}
A x & B x \\
C y & D y
\end{array} \right],
\ee
with $x$ and $y$  the homogeneous coordinates on the two ${\mathbb P}^1$ factors. The bundle ${\bf E}$ is a non-trivial deformation of the tangent bundle (which is the case $B=C=0$ and $A=D= {\bf 1}$).

The corresponding GLSM has a gauge group $U(1)_1\times U(1)_2$, two neutral chiral multiplets $\Sigma_1$, $\Sigma_2$, and the chiral and Fermi pairs $X_I, \Lambda_{I}^X$ ($I=1,2$) and  $Y_{K}, \Lambda_{K}^Y$ ($K=1,2$), with gauge charges $(1,0)$ and $(0,1)$, respectively, and vanishing $R$-charges.  The map \eqref{map def E P1P1} corresponds to  the $\CE$-potentials:
\be
\CE_{ I}^X = \sigma_1 {A_I}^J x_J + \sigma_2  {B_I}^J x_J~, \qquad
\CE_{K}^Y = \sigma_1 {C_K}^L y_L+ \sigma_2  {D_K}^L y_L~,
\ee
with $A, B, C, D$ some generic $2\times 2$ matrices.
We have two $\gamma$-blocks, corresponding to the two ${\mathbb P}^1$ factors, with the corresponding mass matrices:
\be
M_1 = \sigma_1 A+ \sigma_2 B~, \qquad
M_2 = \sigma_1 C+ \sigma_2 D~.
\ee
The application of the residue formula \eqref{JK res formula 22 i} is straightforward. We have $\xi_\eff^\text{UV} \rightarrow +(2, 2) \infty$, so that only the flux sectors with $k_1, k_2\geq 0$ contribute.
The correlation functions are therefore given by a Grothendieck residue:
\be\label{eq:p1p1-corrfn}
\left\langle \CO(\sigma_1,\sigma_2) \right\rangle =
\sum_{k_1, k_2\geq 0} q_1^{k_1} q_2^{k_2}\;
{\rm Res}_{(0)} 
{\CO(\sigma_1,\sigma_2) \, d\sigma_1\wedge d\sigma_2 \ov (\det{M_1})^{k_1+1}  (\det{M_2})^{k_2+1}}~.
\ee
The quantum sheaf cohomology relations of the ${\mathbb P}^1 \times {\mathbb P}^1$ model \cite{McOrist:2008ji,Donagi:2011uz,Donagi:2011va} are given by:
\be\label{qsc P1P1}
\det M_1 \: = \: q_1~, \qquad
\det M_2 \: = \:q_2~.
\ee
These relations can also be read from \eqref{eq:p1p1-corrfn}, since the insertion of $\det E_{1}$ (or $\det E_2$) in the integral is equivalent to shifting $k_1$ (or $k_2$) by one.

The correlation functions \eqref{eq:p1p1-corrfn} can be computed explicitly, for instance by using  standard properties of the residue reviewed in appendix \ref{app: residues}. 
For the two-point functions, one finds:
\be\label{res P1P1 i}
\langle \sigma \sigma \rangle \: = \:
- \alpha^{-1}\, \Gamma_1, \: \: \:
\langle \sigma \widetilde{\sigma} \rangle \: = \: 
\alpha^{-1}\, \Delta, \: \: \:
\langle \widetilde{\sigma} \widetilde{\sigma} \rangle \: = \:
- \alpha^{-1} \,\Gamma_2,
\ee
where we defined
\bea
&\Gamma_1  =  \gamma_{AB} \det D \: - \: \gamma_{CD} \det B~, \qquad
&\Gamma_2  =  \gamma_{CD} \det A \: - \: \gamma_{AB} \det C~,\cr
& \Delta \: = \: (\det A) (\det D) \: - \: (\det B) (\det C)~, \qquad 
&\alpha \: = \: \Delta^2 \: - \: \Gamma_1 \Gamma_2~,
\eea
with
\be\label{def gammaAB CD}
\gamma_{AB} \: = \: \det(A+B) - \det A - \det B, \qquad
\gamma_{CD} \: = \: \det(C+D) - \det C - \det D~.
\ee
One can perform an independent check of this result by 
using \v{C}ech cohomology techniques \cite{qsc-kcs},  as presented in appendix~\ref{ap:ex:p1xp1-cech}, and one finds perfect agreement. The four points functions can be obtained similarly, as discussed in  appendix~\ref{ap:ex:p1xp1-cech}.

It was argued in \cite{Guffin:2007mp} 
that the singular locus of these correlation
functions, {\it i.e.} the locus $\{ \alpha = 0 \}$ in parameter space,
coincides with the locus on which the bundle degenerates. 
This matches expectations from a lore according to which singularities in $\CN{=}(0,2)$ NLSM are determined by singularities in the bundle and not in the base of the target space. 


\subsection{Hirzebruch surface ${\mathbb F}_n$ and orbifold  $\mathbb{WP}^2_{1,1, n}$}
\label{sect:ex:fn}
\begin{table}[t]
\centering
\begin{tabular}{c|cccc}
 &$x_1$  &$x_2$   &  $w$  &$s$    \\
\hline
$U(1)_1$ & $1$ & $1$ & $n$ & $0$ \\
$U(1)_2$ & $0$ & $0$ & $1$ & $1$ \\
\end{tabular}
\caption{Weights of the homogeneous coordinates of ${\mathbb F}_n$. They coordinates $x_I, w, s$ are also the scalar components of the chiral multiplets $X_I, W, S$, and the weights are their gauge charges.}
\label{tab: Fn charges}
\end{table}

The Hirzebruch surface ${\mathbb F}_n$ with $n>0$ can be described in terms of
four homogeneous coordinates $x_I$ ($I=1,2)$, $w$ and $s$, with weights given in Table \ref{tab: Fn charges}.
The deformation ${\bf E}$ of the tangent bundle is described by the
cokernel
\be
0 \: \longrightarrow \: {\cal O}^2 \: \stackrel{*}{\longrightarrow} \:
{\cal O}(1,0)^2 \oplus {\cal O}(n,1) \oplus {\cal O}(0,1) \:
\longrightarrow \: {\bf E} \: \longrightarrow \: 0
\ee
with
\be
* \: = \: \left[ \begin{array}{cc}
A x & Bx \\
\gamma_1\, w + s \,f_n(x) \quad& \beta_1\, w + s\, g_n(x)\\ 
 \gamma_2\, s & \beta_2\, s
\end{array} \right],
\ee
where $A$, $B$ are $2 \times 2$ complex matrices, $\gamma_1, \gamma_2, \beta_1, \beta_2$ are complex constants, and $f_n$, $g_n$ are degree $n$ homogeneous polynomials.
The special case $A = I$, $B = 0$, $f_n = g_n=0$, $\beta_1 = \beta_2=1$,
$\gamma_1 = n$ and
$\gamma_2 = 0$ correspond to the tangent bundle. 

To discuss this class of models, we should distinguish between the two cases $n=1$ and $n\geq 2$. The key difference is that ${\mathbb F}_1$  is a strictly NEF Fano variety (that is, the anti-canonical divisor has a positive intersection with every effective curve). In that case, the RG flow of the K\"ahler class leads to a large volume ${\mathbb F}_1$ geometry in the UV of the NLSM. By contrast, the NLSM on ${\mathbb F}_n$ with $n>1$ would always flow to a singular orbifold  $\mathbb{WP}^2_{1,1, n}$  in the UV. The naive geometric intuition is not reliable in that case, and one should use the orbifold description instead. (See  \cite{Donagi:2014koa} for a similar discussion.)

The  GLSM corresponding to these geometries has a gauge group $U(1)_1\times U(1)_2$ with two neutral chiral multiplets $\Sigma_1, \Sigma_2$, and the chiral and Fermi pairs $X_I, \Lambda^X_I$ $(I=1,2)$, $W, \Lambda^W$ and $S, \Lambda^S$, with gauge charges given in Table \ref{tab: Fn charges} and vanishing $R$-charges. We have the $\CE$-potentials:
\bea
& \CE_I^X = \sigma_1 {A_I}^J x_J + \sigma_2 {B_I}^J x_J~, \cr
&\CE^W = \sigma_1 \left(\gamma_1 w + s f_n(x)\right)+\sigma_2\left(\beta_1w + s g_n(x)\right)~,
\qquad
\CE^S= \sigma_1 \gamma_2 s+ \sigma_2 \beta_2 s~.
\eea
There are three $\gamma$-blocks here, of dimensions $2$, $1$ and $1$ respectively, with Coulomb branch masses:
\be
M_{X} = \sigma_1 A + \sigma_2 B~, \qquad M_W = \sigma_1\gamma_1\: + \: \sigma_2\beta_1~, \qquad
M_S=  \sigma_1\gamma_2 \: + \: \sigma_2 \beta_2~.
\ee
These masses, and the correlators below, are independent of the non-linear deformation encoded in $f(x)$ and $g(x)$, in accordance with the discussion of section \ref{subsec: properties A2 corr}. According to \eqref{JK res formula 22 i}, the correlation functions are given by:
\be \label{eq:fn-corrfn}
\left\langle \CO(\sigma_1, \sigma_2) \right\rangle = \sum_{k_1, k_2\in \Z} q_1^{k_1}q_2^{k_2}\,  {\text{JKG-Res}}\! \left[\eta\right] \, { \CO(\sigma_1, \sigma_2)\;   d\sigma_1\wedge d\sigma_2\ov  (\det M_X)^{1+k_1} (M_W)^{1+ n k_1 + k_2} (M_S)^{1+ k_2}  }~,
\ee
with $\eta=\xi_\eff^\text{UV} \rightarrow +(2+n, 2) \infty$. This is the simplest example of a non-regular JKG residue: depending on the flux sector, there can be up to three divisors intersecting at the origin of $\t \fM\cong \C^2$. Following \eqref{def omegasSP}, we define:
\be
\omega_{Q_X Q_W}= {P_0   \, d\sigma_1\wedge\sigma_2\ov \det M_X \, M_W}~, \qquad
\omega_{Q_X Q_S}= {Q_0  \, d\sigma_1\wedge\sigma_2\ov \det M_X \, M_S}~, \qquad
\omega_{Q_W Q_S}= {  d\sigma_1\wedge\sigma_2\sigma\ov M_W \, M_S}~,
\ee
with $P_0$ and $Q_0$ some generic homogeneous polynomials of degree $1$. Consider first the case of the first Hirzebruch surface ${\mathbb F}_1$. In this $n=1$ case, $\eta=\xi_\eff^\text{UV}$ lies inside the cone generated by $Q_X$ and $Q_W$, which is the `geometric phase' of the GLSM. 
(For any given $n$, both ${\mathbb F}_n$ and   $\mathbb{WP}^2_{1,1, n}$ are classical `phases' of the same GLSM, but only one phase is relevant quantum mechanically.)
Therefore, we  must have:
\bea
&{\text{JKG-Res}}\! \left[\eta\right] \omega_{Q_X Q_W}={\rm Res}_{(0)}   \omega_{Q_X Q_W}~, \quad
&  {\text{JKG-Res}}\! \left[\eta\right] \omega_{Q_X Q_S}={\rm Res}_{(0)}   \omega_{Q_X Q_S}~, \cr
&{\text{JKG-Res}}\! \left[\eta\right] \omega_{Q_W Q_S}=0~.
\eea
One way to describe the corresponding residue is by  first summing the residues in $\sigma_1$ at the roots of $P_X\equiv  \det M_X$, for $\sigma_2$ fixed and generic, before taking the residue at the remaining pole in $\sigma_2$:
\be
{\text{JKG-Res}}\! \left[\eta\right] f(\sigma_1, \sigma_2)\, d\sigma_1\wedge\sigma_2 =
\oint_{\sigma_2=0} {d\sigma_2\ov 2\pi i}\sum_{\sigma_1^{*} || P_X(\sigma_1^*, \sigma_2)=0} \oint_{\sigma_1 = \sigma_1^{*}}    {d\sigma_1\ov 2\pi i} \,f(\sigma_1, \sigma_2) 
\ee
We thus obtain the following  expressions for the two-point functions in this model:
\bea\label{F1 two point fcts explicit}
\langle \sigma_1^2 \rangle & = &&
\tilde{\alpha}^{-1} \left[ \tilde{\Delta} 
\: - \: \beta_1 \beta_2 \det(A+B) \: + \:
(\gamma_1 +\beta_1) (\gamma_2+\beta_2) \det B \right], \cr
\langle \sigma_1 \sigma_2 \rangle & = &&
\tilde{\alpha}^{-1}  \tilde{\Delta}~, \cr
\langle \sigma_2^2 \rangle & = &&
\tilde{\alpha}^{-1} \left[ \tilde{\Delta} \: - \: 
(\gamma_1 + \beta_1) (\gamma_2 + \beta_2) \det A \: + \:
\gamma_1 \gamma_2 \det(A+B) \right]~,
\eea
where we defined
\bea
&\tilde{\Delta} =  \beta_1 \beta_2 \det A \: - \: 
\gamma_1 \gamma_2 \det B~, \cr
&\Phi_i  = \beta_i^2 \det A \: - \: \beta_i \gamma_i \left(
\det(A+B) - \det A - \det B \right) \: + \: \gamma_i^2 \det B~, \cr
&\tilde{\alpha} = 
\Phi_1 \Phi_2~.
\eea
Higher correlation functions can be obtained similarly. 
The JKG residue results match results which were obtained independently through \v{C}ech-cohomology-based arguments, as described explicitly in appendix~\ref{ap:ex:fn-cech}. 

For $n=2$,  $\eta=\xi_\eff^\text{UV}$ lies along the cone boundary $Q_W$ and our residue formula is not valid. For $n>2$,  $\eta=\xi_\eff^\text{UV}$ lies in the cone generated by $Q_W$ and $Q_S$, which correspond to the `orbifold phase'  $\mathbb{WP}^2_{1,1, n}$. The correlation functions can also be obtained in that case, and are to be interpreted in terms of the $\mathbb{WP}^2_{1,1, n}$ geometry. The fact that $\xi_\eff^\text{UV}$ lies outside the geometric phase in FI parameter space translates geometrically to the fact ${\mathbb F}_n$ for $n>2$ is not a NEF  Fano variety \cite{Donagi:2014koa}. For $n>2$, the JKG prescription gives:
\bea
&{\text{JKG-Res}}\! \left[\eta\right] \omega_{Q_X Q_W}=0~, \quad
&  {\text{JKG-Res}}\! \left[\eta\right] \omega_{Q_X Q_S}={\rm Res}_{(0)}   \omega_{Q_X Q_S}~, \cr
&{\text{JKG-Res}}\! \left[\eta\right] \omega_{Q_W Q_S}={\rm Res}_{(0)}   \omega_{Q_W Q_S}~.
\eea
For all values of $n$, the quantum sheaf cohomology ring relations follow  from  \eqref{eq:fn-corrfn}:
\be\label{QSC Fn}
(\det M_X) \, (M_W)^n \: = \: 
q_1~, \qquad
M_W M_S \: = \: q_2~,
\ee
which agrees with \cite{McOrist:2008ji,Donagi:2011uz,Donagi:2011va}.

\subsection{The quintic}
\label{sect:ex:quintic-def}

The quintic Calabi-Yau threefold inside ${\mathbb P}^4$  can be engineered by a $U(1)$ GLSM with a neutral chiral multiplet $\Sigma$,  four chiral and Fermi multiplets $X_i, \Lambda^X_i$ of gauge charge $Q_i=1$ and $R$-charges $r_i=0$, and a chiral and Fermi multiplet pair $P, \Lambda^P$ of gauge charge $Q_p=-5$ and $R$-charge $r_p=2$.%
~\footnote{Note that the non-zero $R$-charge for $P$ means that the corresponding scalar field is twisted, as discussed {\it e.g.} in \cite{gs22,gs02}.}
By a field redefinition, we can take the $\CE$-potentials to be the same as on the $\CN{=}(2,2)$ locus:
\be
\CE_i = \sigma x_i~, \qquad \CE_p= - 5 \sigma p~.
\ee 
The $R$-charge assignment allows to turn on the $J$-potentials:
\be
J_i = p (\d_i G + G_i)~, \qquad J_p = G~, 
\ee
where $G$ is a homogeneous polynomial of degree five in the $x_i$'s and $G_i$ are homogeneous polynomials of degree four. The  condition \eqref{JE constraint} implies:
\be
x^i G_i = 0~.
\ee
The quintic $X$ in  ${\mathbb P}^4$ corresponds to the locus $G=0$, while the polynomials $G_i$ parameterize a deformation ${\bf E}$ of the tangent bundle  $TX$  \cite{Witten:1993yc}.  The $\CN{=}(2,2)$ locus corresponds to $G_i=0$.

As explained in section \ref{subsec: properties A2 corr}, the correlation functions are independent of the $J$-potential, therefore \eqref{JK res formula 22 i} leads to the same results as on the $(2,2)$ locus \cite{Morrison:1994fr,Benini:2015noa, Closset:2015rna}.

\section{Non-abelian examples}
\label{sect:nonabelian}
In this section, we consider some non-abelian GLSMs with an $\CN{=}(2,2)$ locus. We emphasize the case of the Grassmannian with a deformed tangent bundle, whose quantum sheaf cohomology can be studied using our explicit formula for the $A/2$-twisted correlation functions.  A more thorough study of the Grassmannian manifold quantum sheaf cohomology will appear in   \cite{Guo:2015caf, gls2}.

\subsection{Grassmannian manifold with deformed tangent bundle}\label{subsec: Grassmannian}
Consider the Grassmannian manifold ${\rm Gr}(N_c, N_f)$. Its tangent bundle admits $N_f^2-1$ deformations if $1< N_c <N_f-1$.  (If either $N_c=1$ or $N_c=N_f-1$, there are no deformations. One still has a $B$ matrix below but it only describes trivial deformations.)
 The corresponding GLSM contains a $U(N_c)$ vector multiplet, a chiral multiplet $\Sigma$ in the adjoint representation of the gauge group of vanishing $R$-charge, and $N_f$ chiral and Fermi multiplets $\Phi_i, \Lambda_i^\Phi$ ($i=1, \cdots N_f$) in the fundamental representation and with vanishing $R$-charges. 

The most general $\CE$-potential one can write is
\be
\CE^\Phi_i = {A_i}^j \,\sigma \phi_j +  {\rm Tr}(\sigma)\, {B_i}^j \phi_j~,  
\ee 
where in the first term $\sigma$ acts on $\phi_i$ in the fundamental representation, and $A$ and $B$ are generic $N_f\times N_f$ matrices. The $\CN{=}(2,2)$ locus corresponds to $A= {\mathbf 1}$ and $B=0$. We can set $A={\mathbf 1}$ by a field redefinition. The remaining components of $B$ (modulo the trace) correspond to the $N_f^2-1$ deformations of $T{\rm Gr}(N_c, N_f)$.

We have the mass matrices
\be
M_{a} =  \sigma_a \, A +  \left(\sum_{b=1}^{N_c} \sigma_b\right) B~, \qquad a=1, \cdots, N_c~,
\ee
corresponding to the $N_c$ weights of the fundamental representation.  Using \eqref{JK res formula 22 i}, one can write the correlations functions of gauge-invariant polynomials in $\sigma$ as:
\be
\left\langle \CO(\sigma) \right\rangle^{(A/2)}_{\mathbb{P}^1}= \sum_{{\bf k }\in \Z_{\geq 0}} q^{\bf k} \CZ_{\bf k}(\CO)~,
\ee
in terms of the ${\bf k}$-instanton contributions
\be
\CZ_{\bf k}(\CO) = {(-1)^{(N_c-1) {\bf k}}  \over N_c!} \sum_{k_a | \sum_a k_a ={\bf k}}    {\rm Res}_{(0)}  {\prod_{a\neq b} (\sigma_a-\sigma_b)  \ov  \prod_{a=1}^{N_c} (\det{M_a})^{1+ k_a}}\,  \CO(\sigma) \, d\sigma_1\wedge \cdots \wedge d\sigma_{N_c}~,
\ee
where the sum is over partitions of ${\bf k}$ by non-negative integers. Here we used the fact that $\xi_\eff^\text{UV}\rightarrow (1, 1, \cdots, 1)\infty$. The integrand is regular and the JKG residue reduces to the Grothendieck residue in every contributing flux sector.

In the present case, the resummed expression \eqref{JK res formula 22 i TER} is also valid. This gives:
\be\label{Grassmannian resummed residue}
\left\langle \CO(\sigma) \right\rangle^{(A/2)}_{\mathbb{P}^1}=
{1 \over N_c!}\oint_{\p \tfM}
\left( \prod_{a=1}^{N_c}{d\sigma_a \over 2\pi i} \right)
{ \prod_{a\neq b}(\sigma_a-\sigma_b)
\over \prod_{a=1}^{N_c} (\det M_a +(-1)^{N_c} q)} \,  \CO\left(\sigma\right)~.
\ee
This expression makes it obvious that the correlators satisfy the quantum sheaf cohomology relations defined in section \ref{subsec: QSC from CB}. Following that discussion, the QSC relations are satisfied by the solutions to the equations:
\be\label{grass QSC on CB}
 \det M_{a} = (-1)^{N_c-1} q_a~, \quad \forall a~, \qquad\qquad \qquad \sigma_a\neq \sigma_b \quad {\rm if}\quad a\neq b~.
\ee
The expression  \eqref{Grassmannian resummed residue} ensures that the correlation functions satisfy the QSC relations because any insertion of $f(\sigma)$  leads to a vanishing residue, 
\be
\left\langle f(\sigma)\, \CO(\sigma) \right\rangle^{(A/2)}_{\mathbb{P}^1}=0~,
\ee
by the definition of $f(\sigma)$ given in section \ref{subsec: QSC from CB}. The Vandermonde determinant in the numerator of \eqref{Grassmannian resummed residue} imposes the second constraint in \eqref{grass QSC on CB}.

Interpreting these results mathematically goes beyond the scope of this paper. The QSC of the Grassmanian with deformed tangent bundle will be discussed in great detail in \cite{Guo:2015caf, gls2}, where an explicit  gauge-invariant characterization of the ring relations will also be given.

\subsection{Complete intersection Calabi-Yau inside the Grassmannian}

We can similarly describe the correlation functions of Calabi-Yau models engineered by non-abelian GLSMs. Many such $\CN{=}(2,2)$ models have been  introduced in the literature \cite{Hori:2006dk, Jockers:2012zr} and it is straightforward to consider their $\CN{=}(0,2)$ deformations \cite{Jia:2014ffa}.

Consider, for instance, a complete intersection Calabi-Yau (CICY) manifold $X$ inside ${\rm Gr}(N_c, N_f)$ \cite{Hori:2006dk}.  In $\CN{=}(0,2)$ notation, the GLSM consists of a $U(N_c)$ vector multiplet, an adjoint chiral multiplet  $\Sigma$, $N_f$ chiral and Fermi multiplets $\Phi_i$, $\Lambda_i$ ($i=1, \cdots, N_f)$ in the fundamental representation, and $S$ chiral and Fermi multiplets $P_\alpha$, $\Lambda^P_\alpha$ ($\alpha=1, \cdots, S$) in the ${\rm \bf det}^{-Q_\alpha}$ representation of $U(N_c)$. The gauge charges and $R$-charges are summarized in Table \ref{tab: Gr CICY charges}.
Defining the  baryonic fields:
\be\label{def baryons UNc}
B_{i_1 \cdots i_{N_c}} = \epsilon_{a_1 \cdots a_{N_c}} \Phi^{a_1}_{i_1} \cdots \Phi^{a_1}_{i_1}~,
\ee
transforming in the determinant representation of $U(N_c)$,  we consider  $G_\alpha$ a generic homogeneous polynomial of degree $Q_\alpha$ in the baryonic fields \eqref{def baryons UNc}, for each $\alpha=1, \cdots S$. 
On the $\CN{=}(2,2)$ locus, the corresponding $\CE$- and $J$-potentials read:
\bea
&\CE_i^\Phi =  \sigma \phi_i~, \qquad \quad&& J_i^\Phi = \sum_\alpha P_\alpha \d_{\phi_i} G_\alpha~,   \cr
&\CE_\alpha^P =  - Q_\alpha \Tr(\sigma) \,p_\alpha~, \qquad \quad&& J_\alpha^P =  G_\alpha~.
\eea
\label{sect:ex:fn}
\begin{table}[t]
\centering
\begin{tabular}{c|ccccc}
 &$\Sigma$  &$\Phi_i$   &  $\Lambda_i$  &$P_\alpha$ & $\Lambda^P_\alpha$    \\
\hline
$U(N_c)$ & ${\bf 1}$ & ${\bf N_c}$ & ${\bf N_c}$ & ${\rm \bf det}^{-Q_\alpha}$ &${\rm \bf det}^{-Q_\alpha}$ \\
\hline
$U(1)_R$ & $0$ & $0$ & $-1$ & $2$ & $1$ \\
\end{tabular}
\caption{ Gauge representations and $R$-charges in the $A/2$-twisted GLSM  for complete intersection Calabi-Yau manifolds inside ${\rm Gr}(N_c, N_f)$.    }
\label{tab: Gr CICY charges}
\end{table}
The simplest $\CN{=}(0,2)$ deformation we can consider consists in choosing
\be
\CE_i^\Phi =  \sigma \phi_i+   {\rm Tr}(\sigma)\, {B_i}^j \phi_j~,
\ee
while $\CE^P_\alpha$, $J_i^\Phi$ and $J_\alpha^P$ retain their $\CN{=}(2,2)$ form. To preserve supersymmetry, we need to have
\be
{B_i}^j \phi_j   {\d G_\alpha \ov \d \phi_i} =0~, \qquad \quad \forall \alpha~.
\ee
For generic choices of  $G_\alpha$, this is generally impossible unless $B=0$. It might be possible, however, to turn on some $B$-deformations for specific choices of $G_\alpha$. Geometrically,  this would correspond to allowed deformations of the CICY tangent bundle $TX$ at specific higher-codimension loci  in the complex structure moduli space of $X$.

According to \eqref{JK res formula 22 i}, the $A/2$-twisted correlation functions  are given by:
\be\label{correlators_Hori_Tong}
\begin{split}
\langle \CO(\sigma) \rangle^{(A/2)}_{\mathbb{P}^1} &=\frac{(-1)^S}{N!} \sum_{k_a=0}^\infty ((-1)^{N_c-1} q)^{\sum_{a=1}^{N_c} k_a}  
\\
&  {\rm Res}_{(0)}   \, 
 {\prod_{a\neq b} (\sigma_a-\sigma_b)  \prod\limits_{\alpha=1}^S(-Q_\alpha \sum\limits_{a=1}^N \sigma_a)^{1+Q_\alpha \sum_a k_a} \ov  \prod_{a=1}^{N_c} (\det{M_a})^{1+ k_a}} 
~\CO(\sigma)  \, d^{N_c}\sigma~.
\end{split}
\ee
The FI parameter is marginal and can be chosen at will. To obtain \eqref{correlators_Hori_Tong}, we chose  $\eta= \xi_\eff^\text{UV}$  to lie in the geometric phase---see \cite{Closset:2015rna} for a detailed discussion in the $\CN{=}(2,2)$ case, to which \eqref{correlators_Hori_Tong}  reduces if $B=0$.

\section{Generalizations}\label{sec: generalizations}
In this section, we consider two simple generalizations of the results of section \ref{sec: loc in GLSM 1}. The first generalization exists in the presence of a flavor symmetry, in which case one can add ``twisted mass'' deformations similar to the twisted masses that contribute to the central charge on the $(2,2)$ locus. The second generalization is to $B/2$-twisted theories which are related to the $A/2$-twisted theories with a $(2,2)$ locus by a simple dualization procedure  \cite{Sharpe:2006qd}.

\subsection{Masses for the global symmetries}\label{subsec: mass def}
Consider a GLSM with a $(2,2)$ locus that has a flavor symmetry group $\GG^F$, with Lie algebra $\Fg^F$. At a given point in the parameter space spanned by the $\CE_I$-couplings, the global symmetry group will be a subgroup of the symmetry group $\t\GG^F$ of the theory at the $\CN{=}(2,2)$ supersymmetric locus:
\be
\GG^F \subset \t\GG^F
\ee
because the $\CE_I$-couplings transform non-trivially under $\t\GG^F$. The flavor group $\GG^F$ is the subgroup of  $\t\GG^F$ that leaves the $\CE_I$ couplings (and the $J_I$ couplings) invariant. In the case of a geometric target space $X$ with an isometry group $\t\GG^F$, this means that we have a $\GG^F$-equivariant holomorphic bundle over $X$. In the presence of such a global symmetry, one can couple a background vector multiplet in the usual way, with supersymmetric value:
\be
D^F= 2 i f^F_{1\b 1}~.
\ee
 We do not consider any background fluxes for the flavor symmetry in this work, although their inclusion is straightforward.
 
It is natural to introduce a $\Fg^F$-valued background chiral multiplet $\Sigma^F$, with a constant value for the scalar field:
\be
\sigma^F = m^F~.
\ee
This background multiplet  couples to the matter fields through the $\CE_I$-potentials. We must have
\be
\CE_I= \CE_I(\sigma, m^F, \phi)
\ee
some homogeneous polynomials of degree one in $\sigma$, $m^F$. 
On the Coulomb branch, this is:
\be
\CE_I = \sigma_a \, E^a_I(\phi) + {(m^F)_I}^J F_J(\phi)~,
\ee
where $m^F$ transforms in the appropriate representation of $\Fg^F$.  The mass matrix on the Coulomb branch is obtained in the same way as in \eqref{mass matrix MIJ}:
\be
M_{IJ} = \d_J \CE_I  \big|_{\phi=0} =  \sigma_a \, \d_J E^a_I\big|_{\phi=0} + {(m^F)_I}^K \d_J F_K(\phi)  \big|_{\phi=0}~.
\ee
We also define the $\gamma$-blocks as in section \ref{sec: loc in GLSM 1} and the localization argument goes through.
The singularities of the integrand lie are along the divisors
\be\label{divisors mass}
P_{(\gamma, \, \rho_\gamma)}(\sigma, m^F) = \det M_{(\gamma, \, \rho_\gamma)}=0
\ee
in $\t\fM$. 
The correlation functions are given by the JKG residue \eqref{JK res formula 22 i}, with the understanding that  ``$\text{JKG-Res}$'' here stands for the sum of the local JKG residues at all the points in $\t\fM$ where $s\geq \rk$ distinct divisors \eqref{divisors mass} intersect.

\subsubsection{Example: ${\mathbb P}^1 \times {\mathbb P}^1$}
Consider the ${\mathbb P}^1 \times {\mathbb P}^1$ model of section \eqref{sect:ex:p1xp1}. On the $\CN{=}(2,2)$ locus, the theory has a  symmetry group $\t\GG^F= SU(2)\times SU(2)$, which is completely broken for generic values of the constant matrices $A, B, C, D$.  However, if we choose the special locus
\be
C= 0~, \qquad D= {\bf 1}
\ee
in parameter space, we retain a global symmetry $\GG^F= SU(2)$.
The mass matrices are
\be
M_1 = \sigma_1 A+ \sigma_2 B~, \qquad
M_2 =\sigma_2 {\bf 1} + m^F~, \qquad\qquad m^F= \mat{m & 0 \cr 0& -m}~.
\ee
The correlation functions are simply given by:
\be
\left\langle \CO(\sigma_1,\sigma_2) \right\rangle =
\sum_{k_1, k_2\geq 0} q_1^{k_1} q_2^{k_2}\;
{\rm Res}_{(0)} 
{\CO(\sigma_1,\sigma_2) \, d\sigma_1\wedge d\sigma_2 \ov (\det{M_1})^{k_1+1}  (\det{M_2})^{k_2+1}}~,
\ee
where the residue is the global Grothendieck residue (the sum of all the local residues).

\subsection{$B/2$-twisted GLSM from dualization}\label{subsec: halfB dual: loc}
Consider an $\CN{=}(0,2)$ GLSM containing a $\Fg$-valued vector multiplet, a chiral multiplet $P$ in the adjoint representation of $\Fg$, and pairs of chiral and Fermi multiplets $\Phi_i$ and $\Lambda_I$ (with $i=I$) which transform in conjugate representations $\FR_i$ and $\b \FR_i$ of $\Fg$, respectively.

We choose to assign the $R$-charges:
\be
R[P]=0~,\qquad  R[\Phi_i]=r_i ~,\qquad R[\Lambda_i]=-r_i+1~, \qquad r_i \in \Z~,
\ee
which  satisfies the anomaly-free condition \eqref{R nonanomalous}. The corresponding curved-space theory realizes  the so-called $B/2$-twist discussed in \cite{Sharpe:2006qd}. 
The potential functions $\CE_I$ and $J_I$ must have $R$-charges $-r_i+2$ and $r_i$, respectively. 
We choose $\CE_I$ to be independent of $P$ and $J_I$ to be linear in $P$. Classically, this preserves 
the alternative $R$-symmetry:
\be
R_{\h {\rm ax}}[P]=2~, \qquad R_{\h {\rm ax}}[\Phi_i]=0~, \qquad R_{\h {\rm ax}}[\Lambda_i]=-1~. 
\ee
 We would like to compute the correlation functions of the  $B/2$-twisted GLSM on the sphere:
\be\label{CO gen sigma 0 B}
\left\langle \CO(p) \right\rangle^{(B/2)}_{\mathbb{P}^1}~,
\ee
 where $\CO(p)$ is any gauge invariant polynomial in the scalar $p$ of the multiplet $P$, which are operators in the $B/2$-type pseudo-chiral ring  \cite{Sharpe:2006qd}. The presence of the $R_{\h {\rm ax}}$ symmetry leads to simple selections rules for \eqref{CO gen sigma 0 B}. We have the same global anomalies \eqref{Rax gauge anom} and \eqref{Agrav Ahalf} as for the $A/2$-twisted case, with $R_{{\rm ax}}$ replaced by $R_{\h {\rm ax}}$. The correlation functions \eqref{CO gen sigma 0 B} are holomorphic in the various parameters, including the complexified FI parameters.  By the same arguments as in  section \ref{subsec: properties A2 corr}, we also find that \eqref{CO gen sigma 0 B} is independent of the $\CE_I$-couplings and of the non-linear $J_I$-couplings.

This $B/2$-twisted GLSM is related to the $A/2$-twisted GLSM of section \ref{sec: loc in GLSM 1} by identifying $P=\Sigma$ and exchanging the Fermi and anti-Fermi multiplets (this exchanges $\CE_I$ and $J_I$). The two models have isomorphic physics \cite{Sharpe:2006qd}. Interestingly, however, this $B/2$-twisted GLSM does not have a $(2, 2)$ locus. Geometrically, the present class of models correspond an holomorphic bundle ${\bf E}$ over the target space $X$, with ${\bf E}$  a deformation of the cotangent bundle. This is equivalent to the $A/2$-twisted model on the bundle ${\bf E}^*$, with ${\bf E}^*$ being a deformation of the tangent bundle.

The $B/2$-twisted correlation functions \eqref{CO gen sigma 0 B} can be computed  on the ``Coulomb branch'' (with covering space $\t\fM\cong \C^\rk$) spanned by the scalar field $p$ in the chiral multiplet $P$, 
\be
p= (p_a)~, \qquad a=1, \cdots, \rk~.
\ee
The supersymmetric localization argument works similarly to the one in section \ref{sec: loc in GLSM 1}. On $\t \fM$, we have $J_I= p_a E^a_I(\phi)$, the mass matrix is defined by
\be\label{mass matrix hMIJ}
\h M_{IJ} = \d_J J_I  \big|_{\phi=0} =  p_a \, \d_J \h E^a_I\big|_{\phi=0}~,
\ee
and we have the same decomposition in $\gamma$-blocks as before.
We then obtain a result isomorphic to \eqref{JK res formula 22 i}-\eqref{JK res formula 22 ii} for the correlation functions:
\be\label{JK res formula 22 i B}
\left\langle \CO(p) \right\rangle^{(B/2)}_{\mathbb{P}^1}={(-1)^{N_*}\ov |W|} \sum_{k \in \Gamma_{\mathbf{G}^\vee}} q^k\,  {\text{JKG-Res}}\! \left[\eta \right]  \cZ_k^\oneloop(p)\, \CO(p)\,   dp_1\wedge \cdots \wedge dp_\rk ~,
\ee
with
\be\label{JK res formula 22 ii B}
 \cZ_k^\oneloop(p) = 
  (-1)^{\sum_{\alpha>0}(\alpha(k) +1)} \prod_{\alpha>0} \alpha(p)^2 \, \prod_\gamma \prod_{\rho_\gamma \in\FR_\gamma} \left( \det \h M_{(\gamma, \, \rho_\gamma)}\right)^{r_\gamma - 1 - \rho_\gamma(k)}~.
\ee
The notation here is the same as in section \ref{subsec: main result}. The one-loop contribution \eqref{JK res formula 22 ii B} is similar to the $A/2$-twist case, and it is discussed in appendix \ref{subsec: app halfB}.  The formula \eqref{JK res formula 22 i B} can be argued for by using the fact that the $A/2$ and $B/2$ models are isomorphic, with isomorphic supersymmetry transformations after one integrates out the auxiliary fields $\CG_I$ in the Fermi multiplets.

\subsubsection{Example: ${\mathbb P}^1 \times {\mathbb P}^1$ with deformed cotangent bundle}
Consider the $B/2$-twist of the GLSM engineering  the ${\mathbb P}^1 \times {\mathbb P}^1$ model with holomorphic bundle {\bf E} defined by the short exact sequence:
\be
0 \: \longrightarrow \: {\bf E} \: \longrightarrow \:
{\cal O}(-1,0)^2 \oplus {\cal O}(0,-1)^2 \: \stackrel{*}{\longrightarrow}
\: {\cal O}^2 \: \longrightarrow \: 0~,
\ee
with
\be
* \: = \: \left[ \begin{array}{cc}
A x & B x \\
C y & D y \end{array} \right]~,
\ee
which is a deformation of the cotangent bundle of ${\mathbb P}^1 \times {\mathbb P}^1$.

The GLSM consists of a $U(1)_1\times U(1)_2$ vector multiplet, two neutral chiral multiplets $P_1$, $P_2$, the chiral and Fermi multiplets $X_I, \Lambda_{I}^X$ ($I=1,2$) of gauge charges $(1,0)$ and $(-1,0)$, respectively, and $R$-charge $0$, and  the chiral and Fermi multiplets $Y_{K}, \Lambda_{K}^Y$ ($K=1,2$), of gauge charges $(0,1)$ and $(0,-1)$, respectively, and $R$-charge $0$.  The $\CE_I$-potentials vanish and the $J_I$-potentials read:
\be
J_{ I}^X = p_1 {A_I}^J x_J + p_2  {B_I}^J x_J~, \qquad
J_{K}^Y = p_1 {C_K}^L y_L+ p_2  {D_K}^L y_L~,
\ee
with $A, B, C, D$ some constant $2\times 2$ matrices.
The mass matrices are:
\be
\h M_1 = p_1 A+ p_2 B~, \qquad
\h M_2 = p_1 C+ p_2 D~.
\ee
The formula \eqref{JK res formula 22 i B} leads to the Grothendieck residue:
\be
\left\langle \CO(p_1, p_2) \right\rangle =
\sum_{k_1, k_2\geq 0} q_1^{k_1} q_2^{k_2}\;
{\rm Res}_{(0)} 
{\CO(p_1, p_2) \, dp_1\wedge dp_2 \ov (\det{\h M_1})^{k_1+1}  (\det{\h M_2})^{k_2+1}}~,
\ee
which is isomorphic to \eqref{eq:p1p1-corrfn}.

\section{Acknowledgements}

We would like to thank L.~Anderson,
F.~Benini, S.~Cremonesi,  Y.~Deng, Z.~Komargodski, Z.~Lu and I.~Melnikov
for useful discussions.  We thank in particular D.~Park for
many useful discussions and for collaboration at the beginning of this project.
We would also like to thank L.~Anderson for giving us permission to
include here the results of appendix \ref{App D},  which were
originally worked out for \cite{qsc-kcs}.
B.~Jia was partially supported by NSF grant PHY-1316033.
E.~Sharpe was partially supported by NSF grant PHY-1417410.

\appendix

\section{Conventions and review of $\CN{=}(0,2)$ supersymmetry}\label{app: conventions}
\subsection{Curved space conventions}
Our conventions mostly follow \cite{Closset:2014pda,Closset:2015rna}, to which we refer for further details.
We work on a Riemannian two-manifold with local complex coordinates $z, \bz$, and Hermitian metric:
\be
ds^2 = 2 g_{z\b z}(z, \bz) dz d\bz~.
\ee
We choose the canonical frame
\be
e^1 = g^{1\over 4} dz~, \quad e^{\b 1} = g^{1\over 4} d\bz~,
\ee
with $\sqrt{g}= 2 g_{z\bz}$ by definition.  
The spin connection is  given by 
\be
\omega_z=-{i\over 4}\d_z\log g~, \qquad \omega_\bz = {i\over 4}\d_\bz \log g~.
\ee
Our only departure from the conventions of \cite{Closset:2014pda} is that we flip the sign of the Ricci scalar ${\rm R}$, so that ${\rm R}>0$ on the round sphere. 
The covariant derivative on a field of spin $s\in \half \Z$ is:
\be
D_\mu \varphi_{(s)}= (\d_\mu - i s \omega_\mu)\varphi_{(s)}~.
\ee
We generally write down derivatives in the frame basis as well: $D_1 \varphi_{(s)}= e_1^z D_z \varphi_{(s)}$ and $D_{\b 1} \varphi_{(s)}= e_{\b 1}^\bz  D_\bz\varphi_{(s)}$.

\subsection{$\CN{=}(0,2)$ supersymmetry in flat space}\label{app:subsec: susy}
For completeness, let us briefly review $\CN{=}(0,2)$ supersymmetry in flat space, following \cite{Witten:1993yc}. We work in Euclidean signature on $\R^2\cong \C$ in complex coordinates. The $\CN{=}(0,2)$ superspace has coordinates $(z, \b z , \theta^+ , \t \theta^+)$.
The supercharges act on  superspace as:
\be\label{defQp}
Q_+ = {\d \over \d \theta^+} + 2 i \t \theta^+ \d_{\b z} \, ,\quad \qquad \t Q_+ = - {\d \over \d \t \theta^+} - 2 i  \theta^+ \d_{\b z} \, , 
\ee
and satisfy
\be
Q_+^2=0~, \qquad \t Q_+^2 =0~, \qquad \{Q_+, \t Q_+\} = - 4 i \d_{\b z}~.
\ee
The supercovariant derivatives are:
\be
D_+ = {\d \over \d \theta^+} - 2 i \t \theta^+ \d_{\b z}~, \quad\qquad \t D_+ = - {\d \over \d \t \theta^+} + 2 i  \theta^+ \d_{\b z}~,  
\ee
We consider theories with an  $R$-symmetry, $U(1)_R$, which acts on the superspace coordinates with charges $R[\tp]= 1$ and $R[\ttp]=-1$. In the following, we review various supersymmetric multiplet and we briefly discuss their relation to the curved-space twisted multiplets of section \ref{sec: curved space susy}.

\subsubsection{General multiplet} 
The general multiplet $\CS$ corresponds to a superfield
\be
\CS_{(s_0)}= C+ i \tp \chi_+ + i\ttp \t\chi_+ + 2\tp\ttp v_\bz~,
\ee
of spin $s_0$ and $R$-charge $r$. The components 
\be\label{CS in compo}
\CS_{(s_0)}= \left(C~,\, \chi_+~, \,\t\chi_+~, \, v_\bz\right)
\ee
have spin $\left(s_0,s_0-\half, s_0-\half, s_0-1\right)$ and $R$-charge $(r, r-1, r+1, r)$, respectively.
The supersymmetry variations of \eqref{CS in compo} are:
\bea\label{susy flat S}
&\delta C= -i \zeta_- \chi_+ - i \t\zeta_- \t\chi_+~, \cr
&\delta \chi_+= 2i \t\zeta_- \, (v_\bz - i \d_\bz C)~, \cr
&\delta \t\chi_+= -2i \zeta_- \, (v_\bz + i \d_\bz C)~, \cr
&\delta v_\bz = -\zeta_- \d_\bz \chi_+ + \t\zeta_- \d_\bz \t\chi_+~.
\eea
Here $\zeta_-$ and  $\t\zeta_-$ are constant supersymmetry parameters, of $R$-charge $1$ and $-1$, respectively,  and \eqref{susy flat S} realizes the supersymmetry algebra:
\be
\delta_{\zeta}^2 =0~, \qquad \quad \t\delta_{\t\zeta}^2=0~, \qquad \quad \{\delta_{\zeta}, \t\delta_{\t\zeta}\} = -4 i \zeta_- \t\zeta_- \d_\bz~.
\ee
In curved space, we set $\zeta_-=0$ and $\t\zeta_-$ becomes a constant Killing spinor. One can obtain the curved-space multiplet \eqref{gen multiplet Sigma} from flat space by defining fields of vanishing  $R$-charge using $\t\zeta_-$:
\be
{\bf C}= (\t\zeta_-)^r C~, \qquad  \chi_{\b1}=  (\t\zeta_-)^{r-1}\chi_+~, \qquad 
\t\chi=   (\t\zeta_-)^{r+1}\t\chi_+~, \qquad   {\bf v}_{\b1}= (\t\zeta_-)^r v_{\b1}~.
\ee
The curved-space multiplet  \eqref{gen multiplet Sigma} therefore has twisted spin $s= s_0+\half r$.

\subsubsection{Chiral multiplet}
The chiral multiplets $\Phi_i$ and antichiral multiplets $\t\Phi_i$ correspond to general superfields  of spin $s_0=0$ and $R$-charges $r$ and $-r$, constrained by:
\be
\t D_+ \Phi_i=0~, \qquad\qquad D_+ \t\Phi_i=0~.
\ee
In components, 
\be
\Phi_i = \varphi_i + \sqrt{2} \theta^+ \psi_{+ i} - 2 i \theta^+ \t \theta^+ \d_{\b z} \phi_i, \qquad
\qquad\t \Phi_i = \t \varphi_i - \sqrt{2} \t \theta^+ \t \psi_{+i}  +  2 i \theta^+ \t \theta^+ \d_{\b z} \t \phi_i~.
\ee
The fields $(\varphi_i, \psi_{+i})$ have spins $(0,-\half)$ and $R$-charges $(r_i, r_i-1)$, and similarly for the charge conjugate multiplet $\t\Phi_i$. The curved-space twisted fields \eqref{phi tphi components} are defined by
 \bea
&\phi_i = (\t\zeta_-)^{r_i} \varphi_i~,\qquad  && \CC_i=  (\t\zeta_-)^{{r_i}-1} \psi_{+i}~, \cr
& \t\phi_i=(\t\zeta_-)^{-{r_i}} \t\varphi_i~,\qquad  &&\t\CB= (\t\zeta_-)^{-{r_i}+1} \t\psi_{+i}~.
\eea

\subsubsection{Fermi multiplet}
The Fermi multiplet $\Lambda_I$ and the anti-Fermi multiplet  $\t\Lambda_I$ correspond to  general superfields of spin $s_0=\half$ and $R$-charges $r_I$ and $-r_I$, respectively, such that:
\be
\t D_+ \Lambda_I=\sqrt2 E_I~, \qquad\qquad D_+ \t\Lambda_I = -\sqrt2 \t E_I~, 
\ee
where $E_I$ and  $\t E_I$ are themselves chiral and antichiral superfields of $R$-charges $r_I+1$ and $-r_I-1$, respectively, which are given as part of the definition of the Fermi multiplet.  In components, we have
\bea
&\Lambda_I = \lambda_{-I} - \sqrt{2} \theta^+ G_I - 2 i \theta^+ \t\theta^+ \d_{\b z} \lambda_{-I} - \sqrt{2}\t \theta^+ E_I~,\cr
&\t \Lambda_I = \t \lambda_{-I} - \sqrt{2}\t \theta^+ \t G_I + 2 i \theta^+ \t\theta^+ \d_{\b z} \t \lambda_{-I} - \sqrt{2} \theta^+ \t E_I~.
\eea
The fields $(\lambda_{-I}, G_I)$ and $(\t\lambda_{-I}, \t G_I)$ have spin $(\half, 0)$ and $R$-charges $(r_I, r_I-1)$ and $(-r_I, -r_I+1)$, respectively.
 The curved-space twisted fields \eqref{fermi twisted compo i}  and  \eqref{fermi twisted compo ii}   are defined by:
\bea
&\Lambda_I =  (\t\zeta_-)^{r_I}  \lambda_{-I}~, \qquad  &&\CG_I = (\t\zeta_-)^{r_I-1}  G_I~, \qquad &&\CE_I = (\t\zeta_-)^{r_I+1}  E_I~, \cr
& \t\Lambda_I =  (\t\zeta_-)^{-r_I}  \t\lambda_{-I}~, \qquad &&\t\CG_I = (\t\zeta_-)^{-r_I+1} \t G_I~, \qquad &&\t\CE_I = (\t\zeta_-)^{-r_I-1}  \t E_I~.
\eea

\subsubsection{Vector multiplet}
A vector multiplet is a pair $(\CV, \CV_z)$ of  general multiplets of spin $s_0=(0, 1)$ and vanishing $R$-charge, subject to the gauge redundancy \eqref{deltaOmega V}. In WZ gauge, 
the corresponding superfields read:
\be\label{V Vz superfields}
 \CV =   2\tp\ttp a_{\b z}~,
\qquad\quad
 \CV_z =  a_z  +  i \tp  \t\lambda_-  +  i \ttp \lambda_-   - \tp\ttp  D~,
\ee
and the  supersymmetry transformations are given by
\bea
&\delta a_z = - i \zeta_-\t\lambda_- - i \t\zeta_- \lambda_-~, \qquad\qquad &
&\delta a_\bz =0~, \cr
&\delta \lambda_- = i \zeta_- (D+ 2 i f_{z\bz})~, \qquad \qquad&
&\delta \t\lambda_- = -i\t \zeta_- (D- 2 i f_{z\bz})~, \cr
&\delta D= 2 \zeta_- \d_\bz \t\lambda_- - 2 \t\zeta_- \d_\bz \lambda_-~,
\eea
where $f_{z\bz}$ is the field strength
\be
f_{z\bz}= \d_z a_\bz - \d_\bz a_z - i [a_z, a_\bz]~.
\ee
The twisted gaugino in \eqref{vec mult02} are defined by $\t\lambda= (\t\zeta_-)^{-1} \t\lambda_-$ and $\lambda_1= \t\zeta_- \lambda_-$, while $a_\mu$ and $D$ are $R$-neutral and therefore remain untwisted.

\section{Elementary properties of the Grothendieck residue}\label{app: residues}
The Grothendieck residue is defined as follows \cite{Griffiths}.  Let $x=(x_1, \cdots, x_r)$ be  complex coordinates on $\C^r$. Let $f_1(x), \cdots, f_r(x)$ be $r$ distinct functions, holomorphic in a neighborhood of $x=0$,  $U\subset \C^r$, and assume that the $f_i$'s have $x=0$ as a single isolated common zero in $U$. 
The Grothendieck residue is defined on any $(r,0)$-form
\be\label{def omega rform}
\omega = {f_0(x)\ov  f_{1}(x)\cdots  f_{r}(x)} dx_1\wedge\cdots \wedge dx_r~,
\ee
with $f_0$ holomorphic on $U$, as a contour integral
\be\label{def resG app}
{\rm Res}_{(0)} \,  \omega ={1\over (2\pi i)^r} \oint_{\Gamma_{\epsilon}}    \omega~,
\ee
with a real $r$-dimensional contour:
\be
\Gamma_{\epsilon} =\left\{x \in \C^r\,  \big| \, |f_i| =\epsilon_i~, \; i=1, \cdots, r \right\}~,
\ee
oriented by $d(\arg(f_1))\wedge \cdots\wedge d(\arg(f_r))\geq 0$.
This residue is imminently computable. We refer to \cite{Griffiths} for some background on the subject, and to \cite{cattani-dickenstein-sturmfels,tsikh} for some discussions of algorithms for computing the residue in general. 

Here we  summarize two of the most elementary properties of the residue, which are useful in explicit computations.
Let us define the Jacobian determinant
\be
\CJ_f(0) =\det_{ij} {\d f_i \ov \d x_j}(0)~.
\ee
A simple property of the residue is that
\be
 {\rm Res}_{(0)} \,  \omega = {f_0(0)\ov \CJ_f(0)}~,\qquad\qquad {\rm if}\qquad \CJ_f(0) \neq 0~.
\ee
Another interesting property is the transformation law \cite{Griffiths}. Suppose that the two sets of $r$ holomorphic functions on $U$,  $\{f_i\}$ and $\{g_i\}$, both have $x=0$ as isolated common zero, and that there exists an holomorphic matrix $A_{ij}(x)$ such that
\be\label{g eq Af}
g_i = \sum_j A_{i j} f_j~.
\ee
Then, one can prove that:
\be
{\rm Res}_{(0)} \left( {f_0(x) \, dx_1\wedge\cdots \wedge dx_r\ov  f_{1}(x)\cdots  f_{r}(x) } \right) =
{\rm Res}_{(0)} \left( {f_0(x)  \det(A) \, dx_1\wedge\cdots \wedge dx_r\ov  g_{1}(x)\cdots  g_{r}(x) } \right)~.
\ee
One can often compute \eqref{def resG app} by finding an holomorphic matrix $A$ such that the new $\{g_i\}$ defined by \eqref{g eq Af}  are simply given by
\be
g_i = (x_i)^{n_i}~,
\ee
in which case the residue becomes an iterated Cauchy formula:
\be
 {\rm Res}_{(0)} \,  \omega  =\oint {dx_1\ov 2 \pi i} \cdots \oint {dx_r\ov 2 \pi i} { f_0(x)  \det(A) \, dx_1\wedge\cdots \wedge dx_r\ov (x_{1})^{n_1}\cdots  (x_{r})^{n_r}}~.
\ee

\section{One-loop determinants}\label{sec: oneloop det}
Consider the gauge  theories with a $\CN{=}(2,2)$ locus of section \ref{sec: loc in GLSM 1}. In this appendix, we compute the one-loop determinant of the matter fields. The one-loop contribution from the $W$-bosons and their superpartners is exactly the same as in \cite{Closset:2015rna}, to which we refer for further discussions of the gauge sector. We also briefly discuss the one-loop determinants relevant for the $B/2$-twisted models of section \ref{subsec: halfB dual: loc}.

\subsection{Matter determinant for $A/2$-twisted GLSM with $(2,2)$ locus}
The matter sector localization is performed with the kinetic terms of the chiral and Fermi multiplets. Placing oneself at a generic point on the Coulomb branch and expanding the Lagrangian at quadratic order in the matter fields, one finds:
\be\label{Sloc Ahalf matter}
\SL_{\rm loc} = \t\phi^I \Delta_{IJ}^{\rm bos} \phi^J \;+\;  (\t \CB~, \, \t \Lambda)^I \Delta^{\rm fer }_{IJ}  \mat{\Lambda \cr   \CC }^J + i \t \CB^I Q_I(\t\lambda)\phi_I + \half \t\CB^\Sigma_a \t\phi^I (\d_{\t \sigma_a} \t M_{IJ}) \Lambda^J~,
\ee
with the kinetic operators
\be
\Delta_{IJ}^{\rm bos}  = - 4 \delta_{IJ} D_1 D_{\b 1} + \t M_{I K} {M^K}_J + i Q_I(D)~, \qquad
 \Delta^{\rm fer }_{IJ} =  \mat{ \half \t M_{JI} & 2 i D_1 \cr  - 2 i D_{\b 1}  &\; 2 M_{IJ} }~,
\ee
Here $M_{IJ}$ was defined in \eqref{EI pot on CB}, and $Q_I$ are the gauge charges of $\Phi_I, \Lambda_I$. Since the mixing is limited to the $\gamma$-blocks defined in section \ref{subsec: Coulomb branch}, we restrict ourselves to a single block of gauge charge $Q_\gamma$ and effective $R$-charge
\be\label{rgamma def}
{\bf r}_\gamma = r_\gamma - Q_\gamma(k)~,
\ee
in a given flux sector. It is easy to perform the supersymmetric Gaussian integral explicitly. It will be sufficient to focus on the case $\t\lambda= \t\CB^\Sigma=0$. Most modes organize themselves into ``long multiplets''  $(\phi, \t \phi, \Lambda, \CC, \t\CB, \t\Lambda)$ with
\be
- 4 D_1D_{\b1} \phi = \lambda_{(\gamma, k)} \phi~, \qquad  \lambda_{(\gamma, k)} >0~.
\ee
 On the round sphere, we simply have the spectrum:
\be
\lambda_{(\gamma, k)}^{(j)} =  j(j+1) -  {\rs_\gamma\ov 2}({\rs_\gamma\ov 2} -1)~,\qquad j=j_0+ 1, j_0+2~, \cdots~,
\ee
with 
\be
j_0(\rs_\gamma) = {|\rs_\gamma-1|\ov 2} -\half~,
\ee
and each  $\lambda_{(\gamma, k)}^{(j)}$ has multiplicity $2j+1$.  It turns out that we do not need to know the exact spectrum $\{\lambda_{(\gamma, k)}\}$ to carry out the localization argument, therefore the final  result  is valid on any non-degenerate Riemann surface of genus zero. 
The total contribution from the non-zero modes reads:
\be
 Z^{\gamma}_{\rm massive}(\sigma, \t\sigma, \h D)= \prod_{\lambda_{(\gamma, k)}} {\det_\gamma(\lambda_{(\gamma, k)}+ |M_\gamma|^2)\ov \det_\gamma(\lambda_{(\gamma, k)}+ |M_\gamma|^2 + i Q_\gamma (\hD))}~,
\ee 
where $\lambda_{(\gamma, k)}$ runs over the full spectrum of non-zero eigenvalues including their multiplicities, $\det_\gamma$ denotes the determinant in the $\gamma$-block and $ |M_\gamma|^2= \t M_\gamma M_\gamma$.
 The more important contribution comes from the zero-modes, which are of two types depending on ${\bf r}_\gamma$---see \eqref{zero modes count chiral}-\eqref{zero modes count Fermi}. If ${\bf r}_\gamma <1$, there are $|{\bf r}_\gamma-1|$ zero-mode multiplets $(\phi, \t\phi, \Lambda, \t\CB)$ corresponding to $j=j_0({\bf r}_\gamma)$, while if ${\bf r}_\gamma >1$ there are ${\bf r}_\gamma -1$ fermionic zero modes $(\CC, \t\Lambda)$. This gives:
\be
Z^{\gamma}_{\rm zero{\text{-}}modes}(\sigma, \t\sigma, \h D) =\begin{cases}
(\det_\gamma M_\gamma)^{\rs_\gamma -1} &\; {\rm if}\;\;  \rs_\gamma \geq 1~,\cr
\left(\det \b M_\gamma \ov   \det_\gamma\left( |M_\gamma|^2 + i Q_\gamma(\h D)\right) \right)^{1-\rs_\gamma} &\; {\rm if}\;\; \rs_\gamma < 1~.\end{cases}
\ee
The complete one-loop determinant for the matter fields in the $\gamma$-block is therefore
\be
Z^{\gamma}(\sigma, \t\sigma, \h D)= 
 Z^{\gamma}_{\rm massive}(\sigma, \t\sigma, \h D)Z^{\gamma}_{\rm zero{\text{-}}modes}(\sigma, \t\sigma, \h D)~.
\ee
The complete one-loop contribution from the matter fields is obtained by taking the product of such contributions for all the field components $\Phi_{\rho_\gamma}, \Lambda_{\rho_\gamma}$ in the theory.

\subsection{Matter determinant for the $B/2$-twisted model}\label{subsec: app halfB}
Consider the $B/2$-twisted model described in section \ref{subsec: halfB dual: loc}. Setting $\CE_I=0$, the matter sector Lagrangian for a chiral and Fermi multiplet pair of gauge charges $Q_I$ reads:
\be\label{S Bhalf matter}
\SL_{B/2} = \t\phi^I \Delta_{IJ}^{\rm bos} \phi^J \;+\;  (\t \CB~, \,  \Lambda)^I \Delta^{\rm fer }_{IJ}  \mat{\t \Lambda \cr   \CC }^J~,
\ee
with the kinetic operators
\be
\Delta_{IJ}^{\rm bos}  = - 4 \delta_{IJ} D_1 D_{\b 1} + \t {\h M}_{I K} {\h M^K}_J + i Q_I(D)~, \qquad
 \Delta^{\rm fer }_{IJ} =  \mat{ -i \t{\h M}_{JI} & 2 i D_1 \cr  - 2 i D_{\b 1}  &\; i \h M_{IJ} }~,
\ee
Here we considered a given flux sector with a constant background for the $P$ multiplet, and we set the fermionic zero modes to zero. The main difference with \eqref{Sloc Ahalf matter} is that \eqref{S Bhalf matter} is not fully $\delta$-exact, because \eqref{J term lag} is not $\delta$-exact. Moreover, we integrated out $\CG_I$ to arrive at  \eqref{S Bhalf matter}. Nonetheless, we can still carry out the localization argument by some appropriate scaling of the various terms. 

At $\h D=0$, the Gaussian integral with Lagrangian \eqref{S Bhalf matter} only has contributions from the zero modes. Defining $\rs_\gamma$ as in \eqref{rgamma def}, there are  $|{\bf r}_\gamma-1|$ zero-mode multiplets $(\phi, \t\phi, \t\Lambda, \t\CB)$ if $\rs_\gamma <1$ and $\rs_\gamma -1$ fermionic zero modes $(\CC, \Lambda)$ if $\rs_\gamma >1$. This gives the one-loop determinant
\be
Z^{\gamma}_{\rm zero{\text{-}}modes}(p, \t p) =
(\det \h M_\gamma)^{\rs_\gamma -1}~,
\ee
for each $\gamma$-block.

\section{\v{C}ech-cohomology-based results for the correlation functions}\label{App D}
Some of the correlation functions computed in this work can also be obtained independently in the corresponding NLSM, providing us with a non-trivial check of our results.
The NLSM computation is essentially an explicit computation of the relevant sheaf cohomology ring, which can be done  using \v{C}ech-cohomology techniques \cite{Katz:2004nn,Guffin:2007mp, Sharpe:2012kw}. In this appendix, we summarize some results for the ${\mathbb P}^1 \times{\mathbb P}^1$ and ${\mathbb F}_1$ models.
(The computations presented in this appendix were originally worked
out for \cite{qsc-kcs}, and are given here with the permission of  L.~Anderson.)

\subsection{ ${\mathbb P}^1 \times
{\mathbb P}^1$}
\label{ap:ex:p1xp1-cech}
Consider the  $A/2$-twisted ${\mathbb P}^1 \times {\mathbb P}^1$ model of section \ref{sect:ex:p1xp1}.
The idea behind the \v{C}ech cohomology approach is to construct explicit \v{C}ech representatives
of the sheaf cohomology groups and compute their classical cup products
directly.  This was applied in \cite{Katz:2004nn,Guffin:2007mp} to
simpler versions of the ${\mathbb P}^1 \times {\mathbb P}^1$ model.
Recall that a general deformation ${\bf E}$ of the tangent bundle of
${\mathbb P}^1 \times {\mathbb P}^1$  is given by
\be
0 \: \longrightarrow \: {\cal O}^2 \: \stackrel{E}{\longrightarrow} \:
{\cal O}(1,0)^2 \oplus {\cal O}(0,1)^2 \: \longrightarrow \: {\bf E}
\: \longrightarrow \: 0
\ee
where
\be
E \: = \: \left[ \begin{array}{cc}
A x & B x \\
C y & D y
\end{array} \right],
\ee
for $x$ and $y$ the vectors of homogeneous coordinates on the two ${\mathbb P}^1$ factors.
Let us cover ${\mathbb P}^1 \times {\mathbb P}^1$ by open
charts, as
\be
U_{ij} \: = \: \{ x_i \neq 0, y_j \neq 0 \}.
\ee
We then construct representatives of the sheaf cohomology
groups $H^1({\bf E}^*)$.  However,
using the definition above it is straightforward to show that
$H^1({\bf E}^*) \cong H^0({\cal O}^2)$.  Therefore, in order to construct
representatives of the desired sheaf cohomology groups, we can apply
the coboundary map (in the long exact sequence derived from the dual
of the short exact above) to elements of $H^0({\cal O}^2)$.

The first step in the construction of the coboundary map is to lift
elements of $H^0({\cal O}^2)$ to meromorphic sections of
\be
{\cal O}(-1,0)^2 \oplus {\cal O}(0,-1)^2
\ee
In patch $U_{11}$, for example, since both $x_1$ and $y_1$ are nonzero,
the lift should be of the form
\be
L_{11} \: = \: \frac{1}{x_1 y_1} \left[ \begin{array}{c}
a_1 y_1 + b_1 y_2 \\
a_2 y_1 + b_2 y_2 \\
a_3 x_1 + b_3 x_2 \\
a_4 x_1 + b_4 x_2 \end{array} \right]
\ee
for some constants $a_{1 \cdots 4}$, $b_{1 \cdots 4}$.
Then, for example, the lift of $(1,0)^T \in H^0({\cal O}^2)$ is defined
by an $L_{11}$ of the form above such that
\be
E^T L_{11} \: = \: \left[ \begin{array}{c}
1 \\ 0 \end{array} \right]
\ee
Using this constraint, one can solve for the constants $a_{1 \cdots 4}$,
$b_{1 \cdots 4}$.
In particular, there are eight constants ($a_{1\cdots 4}$, $b_{1 \cdots 4}$)
and eight linear equations that
they must satisfy (determined by the coefficients of each $x_i y_j$ in each
of the two entries in the matrix product), so one expects a unique solution.
More generally, it is straightforward to solve for the constants $a_{1\cdots 4}$,
$b_{1 \cdots 4}$ that lift $(1,0)^T$ and $(0,1)^T$ on each
coordinate patch.
The \v{C}ech representatives $Y_{i j, i' j'}$
for lifts on different patches of a given element
of $H^0({\cal O}^2)$ are then determined as differences of the form
\be
Y_{i j, i' j'} \: = \: L_{i' j'} \: - \: L_{i j}
\ee
on the patch $U_{ij} \cap U_{i' j'}$.

At this point, the $Y$'s  give \v{C}ech representatives of a given element
of $H^1({\bf E}^*)$, corresponding to elements of $H^0({\cal O}^2)$.
On ${\mathbb P}^1 \times {\mathbb P}^1$, the cup products of 
pairs of elements of $H^1({\bf E}^*)$ are top-forms, whose
integrals determine
classical (two-point) correlation functions.
In principle, \v{C}ech representatives of those cup products are
formed from the ratio of the minors of a matrix whose columns are
the \v{C}ech representatives above, to the reduced maximal minors of the nullspace
of the map $E$.
The resulting ratios define the cup products of the $Y$'s.

Finally, the two-point correlation functions are in principle determined
as integrals of the form
\be\label{cup prod intetgral}
\langle Y \tilde{Y} \rangle \: = \: \int_{ {\mathbb P}^1 \times
{\mathbb P}^1 } Y \cup \tilde{Y}
\ee
More precisely, in principle the cup product yields an element of
$H^2({\mathbb P}^1 \times {\mathbb P}^1, \wedge^2 {\bf E}^* )$,
so part of the details we are suppressing is the use of the
isomorphism $\det {\bf E}^* \rightarrow K_{ {\mathbb P}^1 \times
{\mathbb P}^1 }$ to get what is honestly a top-form from the cup
product.  (That isomorphism is determined only up to
{\it e.g.} overall phases, and plays an important role when
considering how the correlation functions vary over the moduli space.)
In the language of \v{C}ech cohomology, to explicitly evaluate \eqref{cup prod intetgral} we need a trace that does not see any coboundary that
does not touch every patch, and which extracts pieces proportional to
an inverse power of a product of homogeneous coordinates.  In the
present case, the desired trace has the form
\be
\langle Y \tilde{Y} \rangle \: = \: (x_1 x_2 y_1 y_2) \left(
(Y \cup \tilde{Y})_{1,1; 1,2; 2,1} \: - \:
(Y \cup \tilde{Y})_{1, 2; 2, 1; 2, 2} \right).
\ee
The final result for the two-point functions is given by \eqref{res P1P1 i}, in agreement with the JKG residue formula.
These classical correlation functions obey
\be
\langle \det M_1 \rangle =0~, \qquad
\langle \det M_2 \rangle=0~,
\ee
with $M_1, M_2$ defined in section \eqref{sect:ex:p1xp1}.

To compute the four-point functions, there are two natural approaches.
If one does not know the quantum sheaf cohomology relations,
 the four-point functions can be computed by analogous \v{C}ech methods
on the GLSM moduli spaces \cite{Katz:2004nn,Guffin:2007mp,Sharpe:2012kw}.  Another simpler method is available if we already know the QSC relations, as is the case here
\cite{McOrist:2008ji,Donagi:2011va,Donagi:2011uz}, since one can simply use these relations to derive the four-point functions from the two-point functions  algebraically.
In this case, the QSC relations read:
\be
\det(A \sigma_1 + B \sigma_2) \: = \: q_1, \: \: \:\qquad
\det(C \sigma_2 + D \sigma_2) \: = \: q_2~,
\ee
which gives the following equations for the four-point functions:
\bea
& \langle \sigma_1^4 \rangle \det A + \langle \sigma_1^2 \sigma_2^2 \rangle \det B + \langle \sigma_1^3 \sigma_2 \rangle \gamma_{AB}
& = &\; \;q_1 \,  \langle \sigma_1^2 \rangle~, \cr
& \langle \sigma_1^3 \sigma_2 \rangle \det A + \langle \sigma_1 \sigma_2^3 \rangle \det B + \langle \sigma_1^2 \sigma_2^2 \rangle \gamma_{AB}
& = &\;\;  q_1  \, \langle \sigma_1\sigma_2 \rangle~, \cr
& \langle \sigma_1^2 \sigma_2^2 \rangle \det A + \langle  \sigma_2^4 \rangle \det B + \langle \sigma_1 \sigma_2^3 \rangle \gamma_{AB}
& = &\;\;  q_1  \, \langle \sigma_2^2 \rangle~, \cr
& \langle \sigma_1^4 \rangle \det C + \langle \sigma_1^2 \sigma_2^2 \rangle \det D + \langle \sigma_1^3 \sigma_2 \rangle \gamma_{CD}
& = &\; \;q_2 \,  \langle \sigma_1^2 \rangle~, \cr
& \langle \sigma_1^3 \sigma_2 \rangle \det C + \langle \sigma_1 \sigma_2^3 \rangle \det D + \langle \sigma_1^2 \sigma_2^2 \rangle \gamma_{CD}
& = &\;\;  q_2  \, \langle \sigma_1\sigma_2 \rangle~, \cr
& \langle \sigma_1^2 \sigma_2^2 \rangle \det C + \langle  \sigma_2^4 \rangle \det D + \langle \sigma_1 \sigma_2^3 \rangle \gamma_{CD}
& = &\;\;  q_2  \, \langle \sigma_2^2 \rangle~, \cr
\eea
with $\gamma_{AB}$ and $\gamma_{CD}$ defined in \eqref{def gammaAB CD}.
The resulting expressions agree with the result one can obtain from the residue formula \eqref{eq:p1p1-corrfn}.

\subsection{\v{C}ech-cohomology-based results for ${\mathbb F}_1$}
\label{ap:ex:fn-cech}

\v{C}ech-cohomology-based arguments can also be used to derive the
two-point functions of the ${\mathbb F}_n$ NLSM. In fact, only the ${\mathbb F}_1$ case is relevant for the results of section \ref{sect:ex:fn}, because for $n\geq 2$ the theory does not correspond to the ${\mathbb F}_n$ model in the UV, but to an orbifold phase.

The structure of the \v{C}ech cover
for ${\mathbb F}_1$ is essentially identical to that of 
${\mathbb P}^1 \times {\mathbb P}^1$, therefore the (classical) two-point functions
should be identical, albeit with changes in parameters.  Reading off
results from the ${\mathbb P}^1 \times {\mathbb P}^1$ model and following the
notation of section~\ref{sect:ex:fn}, one recovers \eqref{F1 two point fcts explicit}, in perfect agreement with the residue computation.
Higher-point functions can again be obtained algebraically using the QSC relations \eqref{QSC Fn}.

\bibliographystyle{utphys}
\bibliography{bib2dNeq02}{}

\end{document}